\documentclass[aps,prx,twocolumn,english,superscriptaddress,floatfix,longbibliography,10pt]{revtex4-2}
\usepackage[T1]{fontenc}
\usepackage{xcolor}
\usepackage{graphicx}
\usepackage[english]{babel}
\usepackage{physics}
\usepackage{amsmath,amssymb}
\usepackage{dsfont}
\usepackage{quantikz}
\usepackage[unicode=true,pdfusetitle,
bookmarks=true,bookmarksnumbered=false,bookmarksopen=false,
 breaklinks=true,pdfborder={0 0 0},pdfborderstyle={},backref=false,colorlinks=true]
 {hyperref}

\begin{document}
\title{Computing logical error thresholds with the Pauli Frame Sparse Representation}
\author{Thomas Tuloup}
\email{thomas.tuloup@bull.com}
\affiliation{Eviden Quantum Lab, 78340 Les Clayes-sous-Bois, France}
\author{Thomas Ayral}
 \affiliation{CPHT, CNRS, Ecole Polytechnique, IP Paris, F-91128 Palaiseau, France
 }
 \affiliation{Eviden Quantum Lab, 78340 Les Clayes-sous-Bois, France}

\begin{abstract}
    We introduce a sparse classical representation, a truncation strategy and a shot-efficient sampling method to push the classical prediction of quantum error correction thresholds beyond Clifford operations and Pauli errors.
    As two illustrations of the potential of our method, we first show that coherent noise error thresholds, when computed at the circuit level (i.e taking into account full syndrome circuits) for distances up to $d=9$, are systematically overestimated (by a factor of about 4) by a Pauli-twirling approximation of the noise.
    We then apply our method to the recently introduced magic-state cultivation protocol. We show, through shot-efficient importance sampling, that, at distance $d=5$, the multiplicative factor between the $T$-gate and the $S$-gate injection error rate is not the one conjectured from low-$d$ computations: it can be as large as $7$.
\end{abstract}

\maketitle

\section{Introduction}

Quantum error correction has come to the forefront of quantum computing very recently as the limitations of non-error-corrected processors---the so-called noisy, intermediate-scale quantum (NISQ) processors---have been laid bare by a decade of experimenting with these prototypes, and as first proof-of-principle experiments of quantum error correction have been carried out on various platforms \cite{Ofek2016CatQubits,Arute2019,Chen2021RepetitionCode,Egan2021TrappedIons,Abobeih2022SpinQubits,Krinner2022SurfaceCode,Zhao2022SurfaceCode,Bluvstein2023,Sivak2023,DaSilva2024,Ryan-Andserson2024TrappedIons,Acharya2025,Putterman2025CatQubits,rosenfeld2025magicstatecultivationsuperconducting,Eickbusch2025DynamicSurfaceCode}.

Yet bridging the gap between these early experiments and the large-scale codes needed for useful error correction is a formidable challenge, both from an experimental and theoretical point of view.

In this work, we shall be concerned with the latter and will in particular be interested in the computation of the so-called quantum error correction threshold. This threshold is the minimal level of physical errors required for a given quantum hardware and code to successfully protect quantum information from corruption by the outside world. 
Computing this threshold is very important in practice as it crucially guides the design both of quantum hardware and of quantum codes. And yet it is very difficult to compute.

Typical methods to compute this threshold rely on heavy simplifying assumptions.
The two most widespread assumptions are that the quantum circuits used to perform error correction contain only so-called Clifford operations (e.g Hadamard gates and CNOT gates belong to the Clifford group) and that the noise that afflicts quantum computations is of the Pauli type (namely only $X$, $Y$ and $Z$ errors occur). 
These two assumptions reduce the computational complexity of threshold estimation from exponential to polynomial in the number of qubits (and gates), which allows one to perform large-scale threshold estimations.

However, these assumptions are not fulfilled in general, and even less in current and upcoming generations of quantum processors. For instance, decoherence in quantum processors is not always of the Pauli type. 
Other types of noise are observed in current processors, like amplitude damping (also called relaxation) noise or coherent noise. On the other hand, non-Clifford gates (like T gates, namely $\pi/4$ rotations) are essential to perform universal quantum computations.

The introduction of non-Pauli noise and non-Clifford gates, however, leads to a drastic increase in the computational cost of computing the threshold.
Brute-force methods, with a dense representation of the state of the quantum processor, lead to a price that is exponential in the number of qubits, drastically limiting the code sizes one can explore.
Expansions in T gates using stabilizer rank \cite{Bravyi2016Clifford+T, Bravyi2019simulationofquantum} are exponential in the number of T gates, and limited to non-Clifford gates that are diagonal in the computational basis.
Other, more specific, methods have been used to estimate thresholds. For coherent noise, it is possible to map the surface code to a  Majorana fermion problem~\cite{Bravyi2018coherent, Marton2023coherenterrors}, but this only allows to study noise on a phenomenological level. Tensor-network based simulations~\cite{Darmawan2017, Darmawan2018} have also been developed for arbitrary non-Pauli noise, but are also limited to phenomenological-level noise and assuming perfect measurements.
This lack of methods to compute error thresholds for realistic noise models and quantum circuits typically prevents one from computing accurately the influence of coherent noise on surface codes, with some literature pointing to a negligible influence \cite{Bravyi2018coherent, Greenbaum2018coherent, Huang2019CoherentErrors, Marton2023coherenterrors}, and some other pointing to a larger influence \cite{Magesan2013Coherent, Puzzuoli2014Coherent, Wallman2015Coherent, Sanders2016Coherent, Darmawan2017, Kueng2017Coherent, Darmawan2024, Behrends2024a}.
Likewise, the true error threshold of the recently introduced T-gate cultivation protocol \cite{gidney2024magicstatecultivationgrowing} is still out of reach for relevant sizes, although some recent works based on ZX-calculus \cite{wan2026simulatingmagicstatecultivation, haenel2026tsimfastuniversalsimulator} or on more subtle Monte-Carlo sampling of detector error models \cite{hines2026simulatingquantumerrorcorrection} are showing a lot of promise.

In this work, we propose a new computational method to reliably estimate error thresholds in the presence of non-Pauli noise and non-Clifford errors. It is based on a new compressed representation of the state of the quantum processor, dubbed Pauli Frame Sparse Representation.
It is tailored to quantum error correction: this representation is the most economical for states that are supposed to be preserved by the error correction process, namely the states that are stabilized by the code.

With this method, we show that we can simulate a wide range of quantum error correction protocols including non-Pauli noise and non-Clifford gates, with two prominent examples: the computation of the error threshold for coherent noise with the surface code, and the computation of the logical error rate of the magic stat cultivation protocol~\cite{gidney2024magicstatecultivationgrowing} for $d=5$.

This article is organized as follows. Section~\ref{sec:pauli_frame} develops the Pauli Frame Sparse Representation (PFSR), detailing its construction, operational meaning, and relevance for modeling quantum states that live near a code space. Section~\ref{sec:operations_pfsr} demonstrates how PFSR can be used in practice, by giving update rules for Clifford operators, general unitaries, measurements, and noise channels. Section~\ref{sec:thresholds_pfsr} applies the method to estimate thresholds on the rotated surface code under non-Pauli noise, both on phenomenological- and circuit-level. Section~\ref{sec:cultivation} applies the method to estimate the logical error rate of the magic state cultivation protocol for $d=3$ and $d=5$. Section~\ref{sec:conclusion} concludes with a summary of the main insights and a discussion of promising directions for future investigation.

\section{Pauli Frame Sparse Representation}
\label{sec:pauli_frame}

In quantum error correction, one typically works under the expectation that physically relevant states remain predominantly within—or close to—the so-called code space. This assumption is not merely technical: it reflects the operational viewpoint that the purpose of a code is to carve out a subspace in which logical information is protected, and around which errors act in a controlled and correctable manner. Representing states with respect to this code space therefore provides both conceptual clarity and analytical leverage. It allows us to separate the dynamics into components that preserve logical information and components that drive the state away from the code space, the latter being precisely what the recovery procedure is designed to mitigate. 

Our representation of the quantum state blends stabilizer formalism with a sparse state-vector expansion, all expressed in a Pauli-frame-defined eigenbasis.

\subsection{Stabilizer Frame}

 A general state on $n$ qubits can be decomposed as a linear combination of computational basis states. Storing such a state is very costly due to the size of the computational basis. In our method, instead, we want to decompose our state on a reduced basis that takes advantage of the fact that quantum error correction is going to preserve states living in the so-called code space. More specifically, states of the code space are stabilized by a set of $n$ mutually commuting independent Pauli operators
\begin{equation}
    \mathcal{S} = \{S_0, S_1, ... , S_{n-1} \} \subset \mathcal{P}_n,
\end{equation}
where $\mathcal{P}_n = \{ \pm 1, \pm i \} \times \{ I, X, Y, Z \}^{\otimes n}$ denotes the $n$-qubit Pauli group.
These operators generate an Abelian subgroup $\langle S \rangle \subset \mathcal{P}_n$. We call $\mathcal{S}$ the stabilizer frame. 

$\mathcal{S}$ is a compact way to define a common eigenbasis $\{ \ket{\mathbf{s}}\}_{\mathbf{s} \in \{ 0, 1\}^n}$---called the stabilizer eigenbasis---on which we shall decompose our state.
Here, each label 
\begin{equation}
    \mathbf{s} = (s_0, s_1, ... , s_{n-1}) \in \{ 0, 1\}^n
\end{equation}
encodes the stabilizer eigenvalues according to 
\begin{equation}
    S_i \ket{\mathbf{s}} = (-1)^{s_i} \ket{\mathbf{s}}.
\end{equation}

Thus, the bitstring $\mathbf{s}$ specifies the pattern of stabilizer eigenvalues, and will later be referred to as the label of the basis kets.

In general, the system is initialized in the computational basis state $\ket{0}^{\otimes n}$, we have $S_i = Z_i$, and the initial stabilizer eigenbasis coincides with the computational basis.

\subsection{Sparse vector representation}

Within this stabilizer frame, we express a general, non-stabilizer state as a (typically) sparse superposition
\begin{equation}
    \ket{\Psi} = \sum_{\mathbf{s} \in \mathcal{I}} \alpha_{\mathbf{s}} \ket{\mathbf{s}},
\end{equation}
where $\mathcal{I} \subseteq \{0, 1\}^n$ is a sparse index set of basis labels with nonzero amplitudes $\alpha_{\mathbf{s}} \in \mathbb{C}$.

Thus, at initialization,
\begin{equation}
    \ket{\Psi_0} = \ket{0}^{\otimes n}, \,\ \mathcal{S}_0= \{Z_0, Z_1, ... , Z_{n-1} \}, \,\ \mathcal{I}_0 = \{ \mathbf{0} \} , \,\ \alpha_{\mathbf0} = 1.
\end{equation}

\subsection{Pauli histories and relative phases}

\subsubsection{Pauli history}

It is important to note that a stabilizer basis element $\ket{\mathbf{s}}$ is only defined up to a phase. To avoid errors during our computation, it is necessary to be able to compute the relative phases of two basis elements with the same label.
To do so, we additionally keep track of the so-called Pauli history of each basis element, which is the product of all the Paulis that were applied to go from the reference stabilizer eigenstate ($\ket{\mathbf{0}}$, the +1 eigenstate of all $S_i$) to the current basis ket. Hence, each populated basis element $\ket{\mathbf{s}}$ is associated with its Pauli history $P_{\mathbf{s}}$, with 
\begin{equation}
    \ket{\mathbf{s}} = P_{\mathbf{s}} \ket{\mathbf{0}}
\end{equation}
where $P_{\mathbf{s}} = P_1 P_2 ... P_k$, a product of all the Pauli that were applied to $\ket{\mathbf{0}}$ leading to $\ket{\mathbf{s}}$ being populated.

\subsubsection{Relative phases between two Pauli histories with the same label}
\label{subsubsec:relative_phases}

During simulations, it may happen that contributions with different Pauli histories $P_{\mathbf{s}}^{(1)}$ and $P_{\mathbf{s}}^{(2)}$ end up with the same label $\mathbf{s}$. In this case, we cannot merge them safely the two contributions to the same basis ket $\ket{\mathbf{s}}_1 = P_{\mathbf{s}}^{(1)} \ket{\mathbf{0}}$ and $\ket{\mathbf{s}}_2 = P_{\mathbf{s}}^{(2)} \ket{\mathbf{0}}$ without knowing their relative phase

\begin{equation}
    e^{i \phi_{12}}  = \bra{\mathbf{s}}_1 \ket{\mathbf{s}}_2 = \bra{\mathbf{0}} (P_{\mathbf{s}}^{(1)})^{\dagger} P_{\mathbf{s}}^{(2)} \ket{\mathbf{0}}
\end{equation}

Since $P_{\mathbf{s}}^{(1)}$ and $P_{\mathbf{s}}^{(2)}$ generate the same label, i.e. they have the same commutation/anticommutation relation with each of the $S_i$ of the stabilizer frame. This means the product $(P_{\mathbf{s}}^{(1)})^{\dagger} P_{\mathbf{s}}^{(2)}$ commutes with all the $S_i$ and is therefore generated by the stabilizer frame, so there is a decomposition
\begin{equation}
(P_{\mathbf{s}}^{(1)})^{\dagger} P_{\mathbf{s}}^{(2)}=\gamma\prod_{i\in A} S_i
\end{equation}
for some subset $A\subseteq\{0,\dots,n-1\}$ and some global phase $\gamma\in\{\pm1,\pm i\}$. 

Then, since $\ket{\mathbf{0}}$ is a +1 eigenstate of all the generators $S_i$, we simply have 
\begin{equation}
    e^{i \phi_{12}}  = \bra{\mathbf{0}} \gamma\prod_{i\in A} S_i \ket{\mathbf{0}} = \gamma \braket{\mathbf{0}}{\mathbf{0}} = \gamma.
\end{equation}

\subsection{Summary of the Pauli Frame Sparse Representation}

At any time, the state of the system is represented by a Pauli Frame Sparse Representation (PFSR)

\begin{equation}
    \mathrm{PFSR}(\ket{\Psi)} = (\mathcal{S}, \{ (\mathbf{s}, \alpha_{\mathbf{s}}, P_{\mathbf{s}}) \}_{\mathbf{s} \in \mathcal{I}}),
\end{equation}

where
\begin{itemize}
    \item $\mathcal{S} = \{S_i\}$ defines the stabilizer frame,
    \item $\mathcal{I} \subseteq \{0, 1\}^n$ is the index set of populated basis eigenstates,
    \item $\alpha_{\mathbf{s}} \in \mathbb{C}$ are sparse amplitudes,
    \item $P_{\mathbf{s}}$ tracks Pauli histories for consistent phase bookkeeping.
\end{itemize}

It uniquely describes the state
\begin{equation}
    \ket{\Psi} = \sum_{\mathbf{s} \in \mathcal{I}} \alpha_{\mathbf{s}} \, P_{\mathbf{s}} \ket{\mathbf{0}}. \label{eq:pfsr_psi}
\end{equation}

This representation allows to move away from the stabilizer formalism (pure Clifford regime, where $| \mathcal{I} | = 1$) towards a sparse state-vector simulation of non-Clifford dynamics, while retaining a Pauli-frame structure that enables efficient Clifford updates and error-correction-aware reasoning.

\paragraph{Example.}
For example, let us consider the state $\ket{+} = \frac{\ket{0} + \ket{1}}{\sqrt{2}}$ on one qubit. If we take $\mathcal{S} = \{X\}$ as the stabilizer frame, the state will be represented by the following PFSR:
\begin{itemize}
    \item $\mathcal{S} = \{X\}$
    \item $\mathcal{I} = \{ 0 \}$
    \item $\alpha_{0}  = 1$
    \item $P_{0} = I$.
\end{itemize}
However, we could also choose $\mathcal{S} = \{Z\}$ as our stabilizer frame. In this case, we obtain another PFSR for the same state:
\begin{itemize}
    \item $\mathcal{S} = \{Z\}$
    \item $\mathcal{I} = \{ 0, 1 \}$
    \item $\alpha_{0}  = \frac{1}{\sqrt{2}}$, $\alpha_{1}  = \frac{1}{\sqrt{2}}$
    \item $P_{0} = I$, $P_{1} = X$.
\end{itemize}
Hence the PFSR is not a unique representation, and the choice of the stabilizer frame will have an influence on the sparsity of the vector of amplitudes.

\section{Simulations using Pauli Frame Sparse Representation}
\label{sec:operations_pfsr}

In this section we describe how to update the PFSR to enforce the highest sparsity upon applying quantum operations such as Clifford operators, Pauli operators, generic operators, measurements, and general quantum channels (including coherent noise).

\subsection{Action of Clifford operators on the Pauli Frame Sparse Representation}

Let $C$ be a Clifford operator, i.e.\ an operator such that
\begin{equation}
    C \mathcal{P}_n C^\dagger = \mathcal{P}_n.
\end{equation}

Because Clifford operations map Pauli operators to Pauli operators under conjugation, they act on the stabilizer frame and on the Pauli histories in a closed and efficient way. Note that a specific update rule for Pauli operations will be derived below. It will be useful for computing, among others, scalar products between two PFSRs

\paragraph{Action on a single basis state.}
Consider a stabilizer-basis state
\begin{equation}
\ket{\mathbf{s}} = P_{\mathbf{s}} \ket{\mathbf{0}},
\end{equation}
where $\ket{\mathbf{0}}$ is the reference stabilizer eigenstate ($+1$ eigenstate of all stabilizers $S_i$).  
Applying $C$ gives
\begin{equation}
    C \ket{\mathbf{s}}
    = C P_{\mathbf{s}} \ket{\mathbf{0}}
    = (C P_{\mathbf{s}} C^\dagger) \, C \ket{\mathbf{0}}.
\end{equation}
We now define the new reference eigenstate
\begin{equation}
    \ket{\mathbf{0}}' = C \ket{\mathbf{0}},
\end{equation}
which is stabilized by the updated stabilizer set
\begin{equation}
    \mathcal{S}' = \{ S_i' = C S_i C^\dagger \}_{0 \leq i \leq n-1}
\end{equation}
In this new stabilizer frame,
\begin{equation}
    C \ket{\mathbf{s}} = P'_{\mathbf{s}} \ket{\mathbf{0}}', 
    \qquad P'_{\mathbf{s}} = C P_{\mathbf{s}} C^\dagger.
\end{equation}

\paragraph{Update rule.}
Thus, applying a Clifford operator $C$ to the sparse representation corresponds to:
\begin{itemize}
    \item Stabilizer update: $S_i \;\mapsto\; C S_i C^\dagger$
    \item Pauli-history update: $P_{\mathbf{s}} \;\mapsto\; C P_{\mathbf{s}} C^\dagger$
\end{itemize}
Importantly, this transformation does not alter the amplitude coefficients $\alpha_{\mathbf{s}}$, and does not change the number of nonzero entries in the sparse vector.

\paragraph{Relabeling of basis states.}
Since both the stabilizers and the Pauli histories have changed, the stabilizer eigenvalue pattern (i.e.\ the label $\mathbf{s}$) associated with each component may no longer be accurate in the new frame. To recover the correct labels, one must recompute, for each component, the commutation/anticommutation relations between the updated $P'_{\mathbf{s}}$ and the updated stabilizers $\{S'_i\}$.

In practice, this relabeling step can be deferred for efficiency: successive Clifford operations can be accumulated by composing their conjugations, and the relabeling can be performed only when required (e.g. before applying a non-Clifford operator, measuring an observable, or computing an expectation value).

In the Pauli Frame Sparse Representation, Clifford operations act purely by changing the reference frame: they rotate both the stabilizer generators and all tracked Pauli histories within the Pauli group, while leaving the sparse amplitude vector itself unchanged. This property allows Clifford dynamics to be simulated with negligible computational overhead, postponing some costly updates until a non-Clifford or measurement step is encountered.

This update is illustrated in Fig.~\ref{fig:pfsr_overview} (b).

\paragraph{Example.}
As an example, let's create the Bell pair $\frac{\ket{00} + \ket{11}}{\sqrt{2}}$ in our formalism. Starting with the state $\ket{00}$ which is stabilized by $ZI$ and $IZ$, our PFSR is defined by
\begin{itemize}
    \item $\mathcal{S} = \{ZI, IZ\}$
    \item $\mathcal{I} = \{ 00 \}$
    \item $\alpha_{00}  = 1$
    \item $P_{00} = II$
\end{itemize}
Then, we apply the Clifford gate $H$ on qubit 0. After conjugation of each stabilizer generator and Pauli histories, the PFSR is now
\begin{itemize}
    \item $\mathcal{S} = \{XI, IZ\}$
    \item $\mathcal{I} = \{ 00 \}$
    \item $\alpha_{00}  = 1$
    \item $P_{00} = II$.
\end{itemize}
Finally, we apply the gate $CNOT_{0 \rightarrow 1}$, which after conjugation leaves the PFSR
\begin{itemize}
    \item $\mathcal{S} = \{XX, ZZ\}$
    \item $\mathcal{I} = \{ 00 \}$
    \item $\alpha_{00}  = 1$
    \item $P_{00} = II$.
\end{itemize}
with $|\mathcal{I}| = 1$, this means we are still simply describing the stabilizer state stabilized by $XX$ and $ZZ$. Note that in principle, the label 00 might no longer be accurate, and we might need to compute the commutation/anticommutation relations between $P_{00}$ and the new stabilizer to obtain the corrected labels. However, since here $P_{00}$ is still $II$, the label is correct.

\subsection{Action of non-Clifford operators on the Pauli Frame Sparse Representation}

It is possible to go beyond Clifford operators by applying operations to the sparse vector without touching the stabilizer frame.

\subsubsection{Action of a Pauli operator on the sparse vector}
\label{subsubsec:apply_pauli}

Instead of applying a Pauli to the PFSR as a Clifford operation by updating the stabilizer frame and the Pauli histories, it is possible to leave the stabilizer frame unchanged and instead perform a remapping of the labels of the sparse vector.

Let $\sigma \in \mathcal{P}_n$ be an $n$-qubit Pauli operator, and consider a stabilizer frame 
$\mathcal{S} = \{ S_0, \dots, S_{n-1} \}$ defining the common eigenbasis 
$\{ \ket{\mathbf{s}} \}_{\mathbf{s} \in \{0,1\}^n}$ with $S_i \ket{\mathbf{s}} = (-1)^{s_i} \ket{\mathbf{s}}$.

Each stabilizer generator either commutes or anticommutes with $\sigma$. We define a binary commutation vector 
\begin{equation}
    \label{eq:commuation_vector}
    \mathbf{c}(\sigma) = (c_0, c_1, \dots, c_{n-1}) \in \{0,1\}^n,
\end{equation}

where
\begin{equation}
c_i =
\begin{cases}
0 \,\ \textrm{if}  & [S_i, \sigma] = 0, \\
1 \,\ \textrm{if}  & \{S_i, \sigma\} = 0.
\end{cases}
\end{equation}

\paragraph{Label update.}
Applying $\sigma$ to a stabilizer-basis state $\ket{\mathbf{s}}$ flips exactly those stabilizer eigenvalues corresponding to generators that anticommute with $\sigma$:
\begin{equation}
    \sigma \ket{\mathbf{s}} \;\propto\; \ket{\mathbf{s} \oplus \mathbf{c}(\sigma)},
\end{equation}
where $\oplus$ denotes bitwise XOR.  
Thus, $\sigma$ acts as a deterministic permutation of the basis elements labels.
 
\paragraph{Sparse-vector update rule.}
If the current state is represented as Eq.~ \eqref{eq:pfsr_psi}, then after applying $\sigma$ we obtain
\begin{equation}
\sigma \ket{\Psi}
    = \sum_{\mathbf{s} \in \mathcal{I}} 
        \alpha_{\mathbf{s}} \,
        (\sigma P_{\mathbf{s}}) \ket{\mathbf{0}}
    = \sum_{\mathbf{s} \in \mathcal{I}}
        \alpha_{\mathbf{s}} \,
        P'_{\mathbf{s} \oplus \mathbf{c}(\sigma)} \ket{\mathbf{0}},
\end{equation}
where the new Pauli history for each updated component is
\begin{equation}
    P'_{\mathbf{s} \oplus \mathbf{c}(\sigma)} = \sigma P_{\mathbf{s}}.
\end{equation}

Operationally, this means that applying a Pauli operator never increases the number of nonzero entries in the sparse vector; it merely permutes them and modifies their phases, the latter information being stored in the Pauli history.  

This update is illustrated in Fig.~\ref{fig:pfsr_overview} (c).

\subsubsection{Action of a linear combination of Pauli operators on the sparse vector}

A general operator $O$ acting on $n$ qubits can be expanded in the Pauli basis as
\begin{equation}
    O = \sum_{k} \beta_k \sigma_k, \qquad \sigma_k \in \mathcal{P}_n, \quad \beta_k \in \mathbb{C}.
\end{equation}

\paragraph{Action on a single basis element.}
For any stabilizer-basis state $\ket{\mathbf{s}}$, applying $O$ gives
\begin{equation}
    O \ket{\mathbf{s}} 
        = \sum_k \beta_k \sigma_k \ket{\mathbf{s}} 
        = \sum_k \beta_k \ket{\mathbf{s} \oplus \mathbf{c}(\sigma_k)},
\end{equation}
where each $\sigma_k$ permutes the stabilizer label according to its commutation vector $\mathbf{c}(\sigma_k)$ as defined previously in Eq.~(\ref{eq:commuation_vector}).
Each resulting term inherits an updated Pauli history:
\begin{equation}
    P'_{\mathbf{s} \oplus \mathbf{c}(\sigma_k)} = \sigma_k P_{\mathbf{s}}.
\end{equation}

\paragraph{Action on a general sparse state.}
Given the sparse representation
Eq.~ \eqref{eq:pfsr_psi}, 
the action of $O$ yields
\begin{equation}
    O \ket{\Psi}
    = \sum_{\mathbf{s} \in \mathcal{I}} \sum_k
        \beta_k \, \alpha_{\mathbf{s}} \,
        (\sigma_k P_{\mathbf{s}}) \ket{\mathbf{0}}
    = \sum_{\mathbf{s}, k}
        \beta_k \, \alpha_{\mathbf{s}} \,
        P'_{\mathbf{s} \oplus \mathbf{c}(\sigma_k)} \ket{\mathbf{0}}.
\end{equation}
Operationally, each Pauli component $\sigma_k$ of $O$ produces a new set of contributions obtained by permuting the labels and extending the Pauli histories.

\paragraph{Merging contributions.}
Different terms may produce components with identical basis labels but distinct Pauli histories. These represent the same basis ket but may differ by a relative phase. To maintain a compact sparse representation, all contributions sharing the same label $\mathbf{s}$ are merged as
\begin{equation}
    \alpha_{\mathbf{s}}^{(\mathrm{new})}
    = \sum_{(\mathbf{s}',k):\, \mathbf{s}' \oplus \mathbf{c}(\sigma_k) = \mathbf{s}}
        \beta_k \, \alpha_{\mathbf{s}'} \,
        e^{i \phi(P_{\mathbf{s}}^{\mathrm{ref}}, \sigma_k P_{\mathbf{s}'})},
\end{equation}
where $e^{i \phi(\cdot)}$ denotes the relative phase between different Pauli histories leading to the same label
\begin{equation}
    e^{i \phi(P_1, P_2)}  = \bra{\mathbf{0}} P_1^{\dagger} P_2 \ket{\mathbf{0}},
\end{equation}
that can be computed using the method described in Sec.~\ref{subsubsec:relative_phases}. $P_{\mathbf{s}}^{\mathrm{ref}}$ is simply a reference history that we will keep as the Pauli history for this basis label. The choice of $P_{\mathbf{s}}^{\mathrm{ref}}$ is not important, and we simply pick the Pauli history of the first contribution with that label.

Applying a general operator $O$ therefore consists of:
\begin{enumerate}
    \item For each term $\sigma_k$ in its Pauli expansion, apply the Pauli-update rule to every populated label $\mathbf{s}$;
    \item Reindex the resulting components according to $\mathbf{s} \mapsto \mathbf{s} \oplus \mathbf{c}(\sigma_k)$;
    \item Update Pauli histories as $P'_{\mathbf{s}} = \sigma_k P_{\mathbf{s}}$;
    \item Merge duplicate labels by computing relative phases between Pauli histories and a reference history and summing their amplitudes.
\end{enumerate}

This update is illustrated in Fig.~\ref{fig:pfsr_overview} (d).

Note that in the event that the number of populated basis kets still grows beyond computational tractability, a step of truncation can be added at the cost of some error. This is discussed in Sec.~\ref{sec:truncation}.

\paragraph{Example.} 
Starting from the Bell pair $\ket{\psi} = \frac{\ket{00} + \ket{11}}{\sqrt{2}}$ from the previous example, with PFSR:
\begin{itemize}
    \item $\mathcal{S} = \{XX, ZZ\}$,
    \item $\mathcal{I} = \{ 00 \}$,
    \item $\alpha_{00}  = 1$,
    \item $P_{00} = II$,
\end{itemize}
let us now apply a $T$ gate on qubit 0. It is described by the linear combination of Paulis $T_0 = \cos{\frac{\pi}{8}} \, II - i \sin{\frac{\pi}{8}} \, ZI$. First, we compute the PFSR of $\cos{\frac{\pi}{8}} \, II \ket{\psi}$, which is simply obtained by multiplying coefficients by $\cos{\frac{\pi}{8}}$:
\begin{itemize}
    \item $\mathcal{S} = \{XX, ZZ\}$
    \item $\mathcal{I} = \{ 00 \}$
    \item $\alpha_{00}  = \cos{\frac{\pi}{8}}$
    \item $P_{00} = II$.
\end{itemize}
Then, for the second term, $- i \sin{\frac{\pi}{8}} \, ZI \ket{\psi}$, the Pauli $ZI$ anticommutes with $XX$ and will therefore flip the corresponding label's bit, and we get
\begin{itemize}
    \item $\mathcal{S} = \{XX, ZZ\}$
    \item $\mathcal{I} = \{ 10 \}$
    \item $\alpha_{10}  = -i \sin{\frac{\pi}{8}}$
    \item $P_{10} = ZI$.
\end{itemize}
so the sum of the two gives us a PFSR of  $T_0 \ket{\psi}$ as
\begin{itemize}
    \item $\mathcal{S} = \{XX, ZZ\}$
    \item $\mathcal{I} = \{ 00, 10 \}$
    \item $\alpha_{00}  = \cos{\frac{\pi}{8}}$, $\alpha_{10}  = -i \sin{\frac{\pi}{8}}$
    \item $P_{00} = II$, $P_{10} = ZI$.
\end{itemize}

\subsection{Projective measurements}

We consider projective measurements of an $n$-qubit Pauli operator $P\in\mathcal P_n$.
The two projectors onto the $\pm1$ eigenspaces are
\begin{equation}
    \Pi_{\pm} = \frac{I \pm P}{2}.
\end{equation}
Given a normalized state $\ket{\Psi}$ the outcome probabilities are
\begin{equation}
    p_{\pm} = \bra{\Psi}\Pi_{\pm}\ket{\Psi} = \frac{1\pm\langle P\rangle}{2},\qquad
    \langle P\rangle \equiv \bra{\Psi}P\ket{\Psi}.
\end{equation}
Conditioned on outcome $\pm$ the post-measurement (normalized) state is
\begin{equation}
    \ket{\Psi_{\pm}} = \frac{\Pi_{\pm}\ket{\Psi}}{\sqrt{p_{\pm}}}.
\end{equation}

\paragraph{Computing expectation values.}
To compute the expectation value $\langle P\rangle$, we first compute $\ket{\Psi'} = P \ket{\Psi}$ by applying $P$ to $\ket{\Psi}$ not as a Clifford operation, but as a Pauli operator as described in Sec.~\ref{subsubsec:apply_pauli}, in order to have $\ket{\Psi'}$ and $\ket{\Psi}$ expressed with the same stabilizer frame, i.e. in the same basis. Then we can simply take the scalar product $\braket{\Psi}{\Psi'}$ taking advantage of the orthonormality of the basis, and computing the relative phases as described in Sec.~\ref{subsubsec:relative_phases}.

In the Pauli Frame Sparse Representation there are two qualitatively distinct cases, depending on the commutation pattern of the measured Pauli $P$ with the stabilizer frame $\mathcal S=\{S_i\}$.

\subsubsection{If \texorpdfstring{$P$}{P} commutes with all stabilizer generators}
If $[P,S_i]=0$ for every $i$, then we can decompose it as a product of stabilizers
\begin{equation}
    P = \gamma \prod_{i\in A} S_i
\end{equation}
for some subset $A$ and phase $\gamma\in\{\pm1,\pm i\}$. In particular $P$ is diagonal in the current stabilizer eigenbasis, so each basis ket $\ket{\mathbf{s}}$ is an eigenvector of $P$ with eigenvalue $\pm1$ determined by the subset $A$ and the label $\mathbf{s}$.

Let the current state be as in Eq.~ \eqref{eq:pfsr_psi}.

Let us define the indicator function
\begin{equation}
\chi_{\pm}(\mathbf{s}) = 
\begin{cases}
1 & \text{if } P(P_{\mathbf{s}}\ket{\mathbf{0}})=\pm (P_{\mathbf{s}}\ket{\mathbf{0}}),\\
0 & \text{otherwise},
\end{cases}
\end{equation}
i.e. \(\chi_{\pm}(\mathbf{s})\) selects those labels that lie in the $\pm1$ eigenspace of \(P\). (\(\chi_{\pm}(\mathbf{s})\) can be computed from the parity of label bits in \(A\).)

Then the unnormalized post-measurement vector for outcome $\pm$ is
\begin{equation}
    \Pi_{\pm}\ket{\Psi} = \sum_{\mathbf{s}\in\mathcal I} \alpha_{\mathbf{s}}\,\chi_{\pm}(\mathbf{s})\, P_{\mathbf{s}}\ket{\mathbf{0}}.
\end{equation}

Hence the outcome probabilities and normalized update are
\begin{equation}
p_{\pm} = \sum_{\mathbf{s}\in\mathcal I} |\alpha_{\mathbf{s}}|^2 \,\chi_{\pm}(\mathbf{s}),
\qquad
\ket{\Psi_{\pm}} = \frac{1}{\sqrt{p_{\pm}}}\sum_{\mathbf{s}\in\mathcal I} \alpha_{\mathbf{s}}\,\chi_{\pm}(\mathbf{s})\, P_{\mathbf{s}}\ket{\mathbf{0}}.
\end{equation}

In practice, this means that the projection is applied by simply deleting the labels $\mathbf{s}$ for which $\chi_{\pm}(\mathbf{s}) = 0$, and renormalizing the projected state by multiplying all coefficients by a factor $1 / \sqrt{p_{\pm}}$

\subsubsection{If \texorpdfstring{$P$}{P} anticommutes with at least one stabilizer generator}
\label{subsubsec:anticommuting_proj}

If $P$ anticommutes with one or more stabilizer generators then $P$ is \emph{not} diagonal in the current stabilizer eigenbasis. In this case, simply applying the projector $\Pi_{\pm}$ as an operator would increase the number of populated basis eigenstates, which we wish to avoid. Therefore, we apply $\Pi_{\pm}$ by changing the stabilizer frame, keeping the same number of populated basis eigenstates. This change of stabilizer frame mirrors standard stabilizer-measurement updates but is compatible with the sparse expansion.

\paragraph{New stabilizer frame.}
Pick some generator $S_r$ that anticommutes with $P$ (if there are several, any choice yields an equivalent final stabilizer group up to phases). If more than one generator anticommutes with $P$, update all the other anticommuting generators $S_j$ by multiplying them by $S_r$. This way, all the new stabilizer generators $S_j S_r$ commute with $P$ and $S_r$ is the sole anticommuting generator. Therefore, the new stabilizer frame after projection will be updated by replacing $S_r$ by $P$ (usually, it is replaced by $\pm P$ depending on the measurement outcome, but in practice we will always replace it with $P$ and any extra minus sign will be accounted for in the coefficients $\alpha_{\mathbf{s}}$).

Hence the update rule for the stabilizer frame is as follow:
\begin{equation}
    \begin{aligned}
        S'_{i \neq r} & \leftarrow \begin{cases}
            S_i \,\ \text{if} \,\ [S_i, P] = 0 \\
            S_i S_r \,\ \text{if} \,\ \{S_i, P\} = 0
        \end{cases} \\
        S'_r & \leftarrow P
    \end{aligned}
\end{equation}

\paragraph{New Pauli histories.}
In order to update the Pauli histories of all the populated basis eigenstates, let us make the following observation: in the new stabilizer frame we have ${n-1}$ generators $S'_i$ with $i \neq r$, that all commute with each other, and commute with both $S_r$ and $P$. Meanwhile, $S_r$ and $P$ anticommute with each other. Hence, there exists a Clifford operator $U$ that maps each $S'_i$ to a $Z_j$ with $0 \leq j \leq n-1$, maps $S_r$ to $X_n$ and $P$ to $Z_n$.

Then for a given Pauli history $P_{\mathbf{s}}$, the new Pauli history is given by 
\begin{equation}
\label{eq:history_anticom_proj}
    Q_{\mathbf{s}} = \begin{cases}
        \alpha U^{\dagger} (P_{\textrm{else}} \otimes I) U \,\ \text{if $\Pi_+$ is applied} \\
        \beta U^{\dagger} (P_{\textrm{else}} \otimes X) U \,\ \text{if $\Pi_-$ is applied} 
    \end{cases}
\end{equation}

where $P_{\textrm{else}} \in \mathcal{P}_{n-1}$ is the Pauli applied on the first $n-1$ qubits from the conjugation of $P_{\mathbf{s}}$ by $U$

\begin{equation}
    U P_{\mathbf{s}} U^{\dagger} =  \underbrace{P_{\textrm{else}}}_{n-1 \,\ \text{qubits}} \otimes P_n
\end{equation}

and $\alpha, \beta$ are the coefficients of 

\begin{equation}
    P_n \ket{+} = \alpha \ket{0} + \beta \ket{1}
\end{equation}

the recipe to find $U$ and the proof of Eq.~(\ref{eq:history_anticom_proj}) can be found in Appendix~\ref{apdx:anticom_measurement}.

Finally, once the histories have been updated, one needs to renormalize the coefficients of the populated basis kets by multiplying them by a factor $1 / \sqrt{p_{\pm}}$. 

Note that since we changed the stabilizer frame, the labels of the basis kets after the projection will no longer be up-to-date, just like after the application of a Clifford operator. 

\subsubsection{Summary of projective measurement}

Therefore, the application of the projective measurement of a Pauli $P$ is done as follows:
\begin{enumerate}
    \item Compute the probability of measuring a given eigenvalue $p_{\pm}$, and determine which result is observed
    \item Check the commutation/anticommutation of each generator of the stabilizer frame $S_i$ with $P$
    \item Depending on the relations:
    \begin{itemize}
        \item If all generators commute, do not change the stabilizer frame, just delete the labels with the wrong eigenvalue and renormalize the coefficients of each populated basis ket
        \item If at least one generator anticommutes with $P$:
        \begin{enumerate}
            \item Update stabilizer frame
            \item Compute the Clifford $U$, the Pauli $P_{\textrm{else}}$ and the coefficient $\alpha$ or $\beta$
            \item Update the Pauli histories depending on the measurement result
            \item Renormalize the coefficients of each populated basis ket.
        \end{enumerate}
    \end{itemize}
\end{enumerate}

\begin{figure*}[t]
    \centering
    \includegraphics[width=\textwidth]{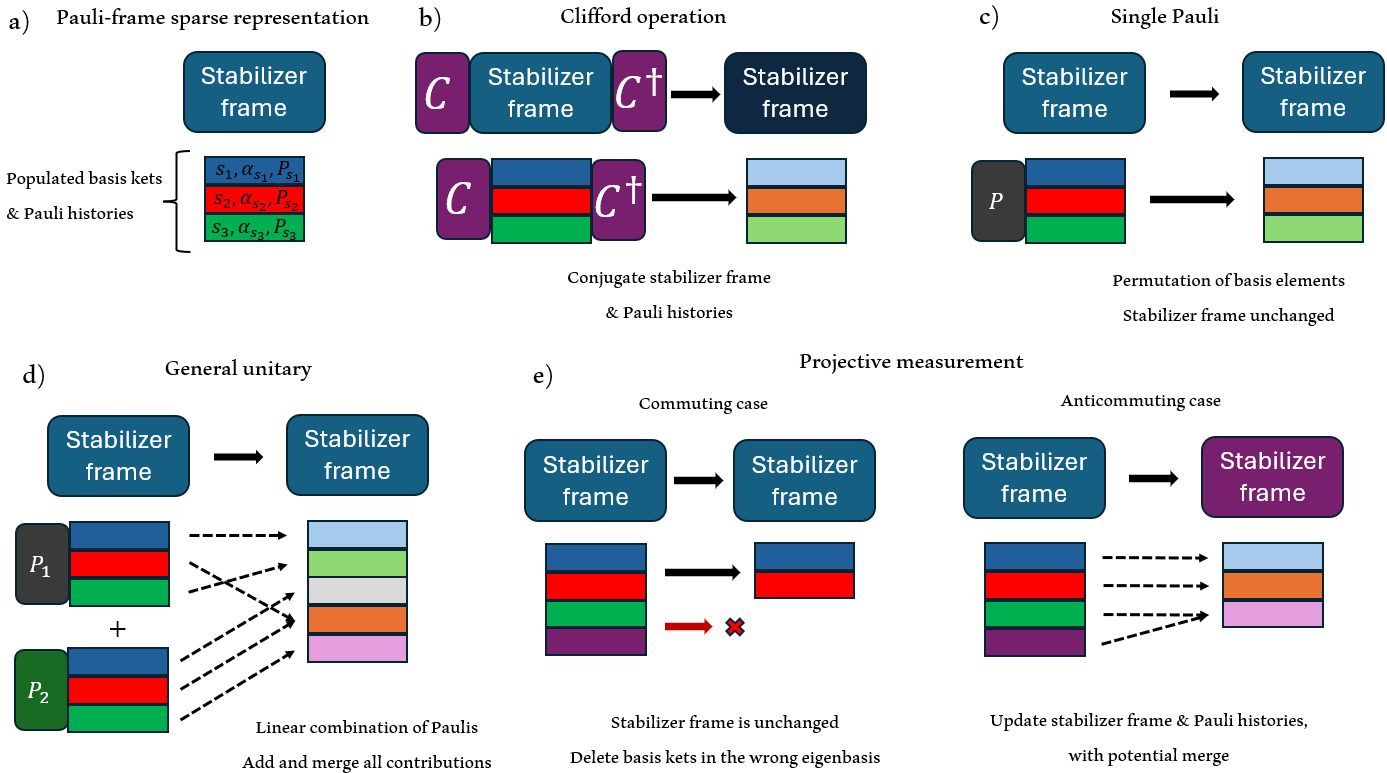}
    \caption{{\it Overview of the Pauli Frame Sparse Representation, and how different operations affect the different components}. a) an instance of PFSR is composed of a stabilizer frame, and a sparse vector of populated basis kets, with amplitude, labels, and Pauli histories. b) When applying a Clifford operator, one may simply conjugate both the stabilizer frame and the Pauli histories, leaving the size of the sparse vector unchanged. c) When applying a single Pauli operator, one may also choose to leave the stabilizer frame unchanged by composing the Pauli histories and relabelling accordingly. d) Any general unitary can be decomposed as a linear combination of Paulis that can be applied to the basis kets individually, before adding and merging the basis kets. In the worst case scenario, this will multiply the size of the vector by the number of Paulis in the linear combination. e) When performing a projective measurement, we distinguish two cases: if the measured Pauli commutes with all stabilizers of the stabilizer frame, the latter does not need to be updated, and we can simply perform the projection by deleting the basis kets that are outside of the eigenspace we project to. This will on average divide the size of the sparse vector by two. If the measured Pauli anticommutes with at least one stabilizer of the stabilizer frame, one must update the stabilizer frame and the Pauli histories according to the method described in Sec.~\ref{subsubsec:anticommuting_proj}. The size of the sparse vector will usually stay the same, up to merging of some Pauli histories.}
    \label{fig:pfsr_overview}
\end{figure*}

This update is illustrated in Fig.~\ref{fig:pfsr_overview} (e).

\paragraph{Example.} 
Starting from the state $\ket{\phi} = T_0 \ket{\psi}$ of our previous example, with PFSR:
\begin{itemize}
    \item $\mathcal{S} = \{XX, ZZ\}$
    \item $\mathcal{I} = \{ 00, 10 \}$
    \item $\alpha_{00}  = \cos{\frac{\pi}{8}}$, $\alpha_{10}  = -i \sin{\frac{\pi}{8}}$
    \item $P_{00} = II$, $P_{10} = ZI$,
\end{itemize}
we now decide to measure qubit 0 in the computational basis, i.e. measure the Pauli $ZI$. First, we evaluate the expectation value $\bra{\phi} ZI \ket{\phi}$. We obtain a PFSR of $ZI \ket{\phi}$ by applying $ZI$ as a Pauli operator and not as a Clifford operator, to leave the stabilizer basis unchanged. We get the PFSR
\begin{itemize}
    \item $\mathcal{S} = \{XX, ZZ\}$
    \item $\mathcal{I} = \{ 00, 10 \}$
    \item $\alpha_{00}  = -i \sin{\frac{\pi}{8}}$, $\alpha_{10}  = \cos{\frac{\pi}{8}}$
    \item $P_{00} = II$, $P_{10} = ZI$.
\end{itemize}
Hence, we obtain $\bra{\phi} ZI \ket{\phi} = -i \cos{\frac{\pi}{8}} \sin{\frac{\pi}{8}} \bra{00} \ket{00} +  i \cos{\frac{\pi}{8}} \sin{\frac{\pi}{8}} (\bra{00} ZI) (ZI \ket{00}) = 0$, so the probability of measuring +1 or -1 is $\frac{1}{2}$.

Let us assume for this example that the measurement gives the result -1. By checking the commutation/anticommutation relations of $ZI$ with $\mathcal{S}$, we notice that $ZI$ anticommutes with $XX$ and commutes with $ZZ$. In order to apply the projection, we first need to find a Clifford $U$ that maps $ZZ$ to $ZI$, $XX$ to $IX$, and $ZI$ to $IZ$. Such a Clifford is given by $U = CNOT_{0 \rightarrow 1} CNOT_{1 \rightarrow 0}$. We can then use this $U$ to update the Pauli histories. 

For $P_{00} = II$, we have $U P_{00} U^{\dagger} = II$ so $P_{\textrm{else}} = I$ and $P_n = I$ so $\beta = \frac{1}{\sqrt{2}}$, and $Q_{00} = \beta U^{\dagger} (P_{\textrm{else}} \otimes X) U = \frac{1}{\sqrt{2}} U^{\dagger} (I \otimes X) U = \frac{1}{\sqrt{2}} XX$. 

For $P_{10} = ZI$, we have $U P_{10} U^{\dagger} = IZ$ so $P_{\textrm{else}} = I$ and $P_n = Z$ so $\beta = -\frac{1}{\sqrt{2}}$, and $Q_{10} = \beta U^{\dagger} (P_{\textrm{else}} \otimes X) U = -\frac{1}{\sqrt{2}} U^{\dagger} (I \otimes X) U = -\frac{1}{\sqrt{2}} XX$. 

We can notice that both $P_{00}$ and $P_{10}$ are being sent to the same Pauli history up to a phase. We can therefore merge those by summing their coefficients, which become (accounting for the renormalizing factor $1 /\sqrt{p_-} = \sqrt{2}$) $\alpha = \sqrt{2} (\frac{1}{\sqrt{2}} \cos{\frac{\pi}{8}} - \frac{1}{\sqrt{2}} (-i) \sin{\frac{\pi}{8}} ) = e^{i \frac{\pi}{8}}$

Since the updated stabilizer frame is now $\mathcal{S} = \{ZI, ZZ\}$ (we replaced the anticommuting $XX$ by the measured Pauli $ZI$), the label of this new Pauli history $XX$ is $10$ and the PFSR after projective measurement is:
\begin{itemize}
    \item $\mathcal{S} = \{ZI, ZZ\}$
    \item $\mathcal{I} = \{ 10 \}$
    \item $\alpha_{10}  = e^{i \frac{\pi}{8}}$
    \item $P_{10} = XX$.
\end{itemize}

\subsection{Application of noise channels}

Since our simulation is based on a sparse representation of the state vector and not the density matrix, noise channels will be applied in a stochastic way, by randomly picking one of the Kraus operators to apply to the state vector at each application of a noise channel, and averaging the result by Monte-Carlo sampling over a large number of trajectories. Here we present the noise channels studied in this work.

\subsubsection{Depolarizing noise}

A depolarizing noise channel with parameter $p$ acts on the density matrix as 

\begin{equation}
    \mathcal{E}(\rho) = (1-p) \rho + \frac{p}{3} X \rho X +  \frac{p}{3} Y \rho Y + \frac{p}{3} Z \rho Z.
\end{equation}

Hence, when applying a depolarizing noise channel with parameter $p$ to the state vector in a stochastic way, we apply either $I$, $X$, $Y$ or $Z$ with respective probabilities $1-p$, $\frac{p}{3}$, $\frac{p}{3}$ and $\frac{p}{3}$. 

Note that since all the operators we can apply are Clifford operators, we could very well apply them by updating the stabilizer frame. However, as a design decision to separate errors from other Clifford operations and to maintain coherency with the other types of noise channels, we apply them instead by updating the labels of the sparse vector (see Fig.~\ref{fig:pfsr_overview} (c)). In either case, all Pauli noises keep the number of populated basis kets constant.

\subsubsection{Amplitude damping noise}
\label{subsubsec:amplitude_damping}

An amplitude damping noise channel with parameter $\gamma$ acts on the density matrix as 

\begin{equation}
    \mathcal{E}(\rho) = K_0 \rho K_0^{\dagger} + K_1 \rho K_1^{\dagger},
\end{equation}
where
\begin{equation}
    \begin{aligned}
        K_0 &= \begin{pmatrix}
            1 & 0 \\ 0 & \sqrt{1 - \gamma}
        \end{pmatrix} = \frac{1 + \sqrt{1 - \gamma}}{2} I + \frac{1 - \sqrt{1 - \gamma}}{2} Z, \\
        K_1 &= \begin{pmatrix}
            0 & \sqrt{\gamma} \\ 0 & 0
        \end{pmatrix} = \frac{\sqrt{\gamma}}{2} X + i \frac{\sqrt{\gamma}}{2} Y.
    \end{aligned}
\end{equation}

Note that unlike Pauli noise channels, the Kraus operators $K_0$ and $K_1$ are non-unitary, and hence do not preserve the norm of the state vector. In this case, in order to stochastically apply this noise channel to a state vector $\ket{\psi}$, we must proceed as follows:

\begin{itemize}
    \item for all Kraus operator $K_i$, compute $\ket{\tilde{\psi_i}} = K_i \ket{\psi}$. $\ket{\tilde{\psi}_i}$ is not a normalized state.
    The probability of applying $K_i$ is then given by $p_i = \braket{\tilde{\psi}_i}{\tilde{\psi}_i}$;
    \item Draw a Kraus operator $K_i$ from this probability distribution, apply it to the state vector and renormalize it, so that the final state after application is $\ket{\psi_i} = \frac{\ket{\tilde{\psi}_i}}{\sqrt{\braket{\tilde{\psi}_i}{\tilde{\psi}_i}}}$.
\end{itemize}

Naturally, when we only have two Kraus operators like in the case of amplitude damping, we simply compute $p_0$, and either apply $K_0$ or $K_1$ with respective probabilities $p_0$ and $1-p_0$.

Note that both $K_0$ and $K_1$ are linear combinations of two Paulis. This means that no matter which one is drawn, the number of populated basis kets might increase up to a factor 2. This can be very problematic when applying a wall of noise on each qubit of a large code, as the vector would quickly lose its sparsity. We present two ways to circumvent this issue. The first one relies on a layering of noise applications (increasing the size of the vector) and projective measurement of stabilizers (reducing the size of the vector) and is presented in Sec.~\ref{subsubsec:layered_noise}. The second one relies on a stabilizer channel decomposition~\cite{Bennink2017} of the amplitude damping channel, and is presented in Appendix~\ref{apdx:stabilizer_channel_decomposition}. In the rest of this work, we will use the first option.

\subsubsection{Coherent noise}
\label{subsubsec:coherent_noise}

Rather than a probabilistic process where errors act randomly on subsets of qubits, noise in a realistic device will often be coherent, i.e., unitary, and can involve small rotations acting everywhere. 

In this work, the coherent noise we will study is a rotation along the Z-axis of a small angle $\theta$

\begin{equation}
    \mathcal{E}(\rho) = e^{-i\frac{\theta}{2} Z} \rho e^{i\frac{\theta}{2} Z},
\end{equation}

Since $e^{-i\frac{\theta}{2} Z} = \cos{\frac{\theta}{2}} I - i \sin{\frac{\theta}{2}} Z$, applying the rotation as is will again increase the number of populated basis kets up to a factor 2. Coherent noise channels also have a stabilizer channel decomposition, which is given in Appendix~\ref{apdx:stabilizer_channel_decomposition}.

\section{Computing thresholds on the rotated surface code}
\label{sec:thresholds_pfsr}

We focus here on the rotated surface code~\cite{Dennis2002SurfaceCode, Kitaev2003, Wang2011RotatedSurfaceCode, Folwer2012SurfaceCode, Fowler2015SurfaceCode} because its local stabilizer structure makes it directly implementable on present-day architectures, and it has been used extensively in experimental demonstrations of error correction. This section applies our PFSR framework to this setting, allowing us to study how a more faithful representation of realistic noise alters logical error rates and threshold behavior.

\subsection{The rotated surface code}

In this work, we will use the $[L^2, 1, L]$ rotated surface code~\cite{Wang2011RotatedSurfaceCode, Tomita2014RotatedSurfaceCode, Litinski2019gameofsurfacecodes}. It is a topological quantum error-correcting code defined on a $L \times L$ square lattice of data qubits with alternating plaquette stabilizers of $X$-type and $Z$-type. Compared with the standard surface code layout, the rotated construction uses fewer physical qubits for a given code distance while preserving locality and the same error-correction properties~\cite{orourke2024comparepairrotatedvs}. 

\paragraph{Lattice structure.}
The code is defined on a two-dimensional square lattice whose faces alternate between $X$- and $Z$-stabilizers. Each face $f$ corresponds to a stabilizer generator
\begin{equation}
    S_f =
\begin{cases}
\displaystyle \prod_{i\in f} X_i, & f \text{ an $X$-type face},\\[6pt]
\displaystyle \prod_{i\in f} Z_i, & f \text{ a $Z$-type face},
\end{cases}
\end{equation}
where the product runs over the data qubits $i$ located at the vertices of the face. Additionally, the rotated surface code uses “rough” and “smooth” boundaries corresponding respectively to $Z$- and $X$-type edges, hosting weight 2 stabilizers.

\begin{figure}[ht]
    \centering
  \includegraphics[width=0.95\linewidth]{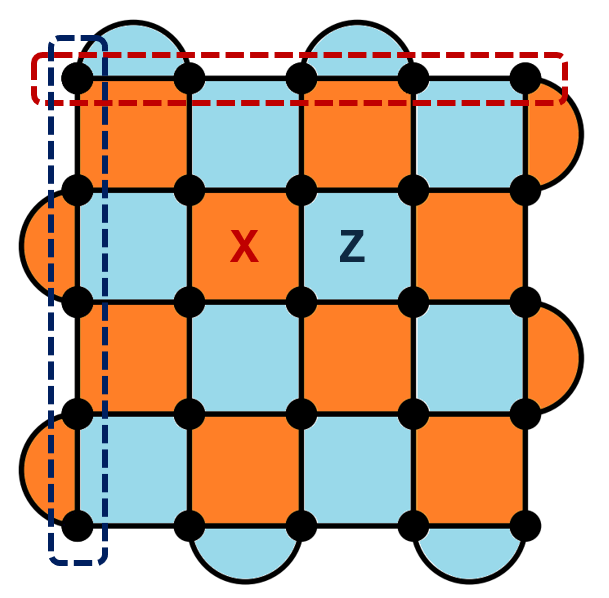}
  \caption{Patch of rotated surface code with distance $d=5$. Black dots represent physical data qubits, and orange (blue) faces represent $X$-type ($Z$-type) stabilizers. The dotted red (blue) contours represent logical operators $X_\mathrm{L}$ ($Z_\mathrm{L})$. Each plaquette and each edge hosts an additional ancilla qubit (not shown) used to measure the corresponding stabilizer.}
  \label{fig:basic_surface_d5}
\end{figure}

The stabilizers mutually commute and generate the stabilizer group $\mathcal{S} = \langle S_f \rangle$. The logical subspace is the simultaneous $+1$ eigenspace of all $S_f$.

For a code of distance $d$, the number of data qubits is $n = d^2$, and the number of ancilla qubits is $d^2 - 1$, so there are $2 d^2 - 1$ physical qubits per logical qubits.
Two logical operators can be chosen as string operators
\begin{equation}
    X_\mathrm{L} = \prod_{i\in \Gamma_X} X_i, \qquad
    Z_\mathrm{L} = \prod_{i\in \Gamma_Z} Z_i,
\end{equation}
where $\Gamma_X$ and $\Gamma_Z$ are nontrivial homologically distinct paths connecting opposite boundaries of the lattice.

\paragraph{Syndrome extraction.}
In a full QEC cycle, each stabilizer generator is measured by an ancilla qubit coupled locally to the four (or two, on edges) data qubits defining that stabilizer. The measurement outcomes $\{s_f = \pm 1\}$ constitute the syndrome, which identifies the pattern of stabilizer violations caused by physical errors. 

In the context of this work, we focus on the quantum memory experiment, where the logical qubit is initialized either in the logical $\ket{0}_\mathrm{L}$ state, corresponding to the $+1$ eigenstate of all $Z$-type stabilizers and of the logical operator $Z_\mathrm{L}$,  or in the logical $\ket{+}_\mathrm{L}$ state, corresponding to the $+1$ eigenstate of all $X$-type stabilizers and of the logical operator $X_\mathrm{L}$. It is then subjected to noise for $d$ rounds + one last round of perfect measurements, and subsequently decoded to determine whether a logical error occurred.

\subsection{Threshold with phenomenological-level noise}

In order to benchmark quantum error correction independently of the details of syndrome extraction circuits, we first estimate thresholds under a phenomenological-level noise model. This approach introduces errors directly on data qubits and stabilizer measurements, rather than modeling the full quantum circuit implementing each stabilizer check. It therefore captures the essential fault-tolerance behavior of the code while avoiding the overhead of simulating ancilla qubits, entangling gates, and measurement operations at the circuit level.

\subsubsection{Layered approach to noise application}
\label{subsubsec:layered_noise}

In our simulations, we employ the phenomenological model to study the intrinsic performance of the rotated surface code under non-Pauli noise, using the Pauli Frame Sparse Representation described in Sec.~\ref{sec:pauli_frame}. At each QEC round, the physical noise channel is applied independently to every data qubit. Syndrome extraction is modeled ideally except for the possibility of independent bit-flip errors on the measured stabilizer outcomes.

\begin{figure}[ht]
    \centering
  \includegraphics[width=0.95\linewidth]{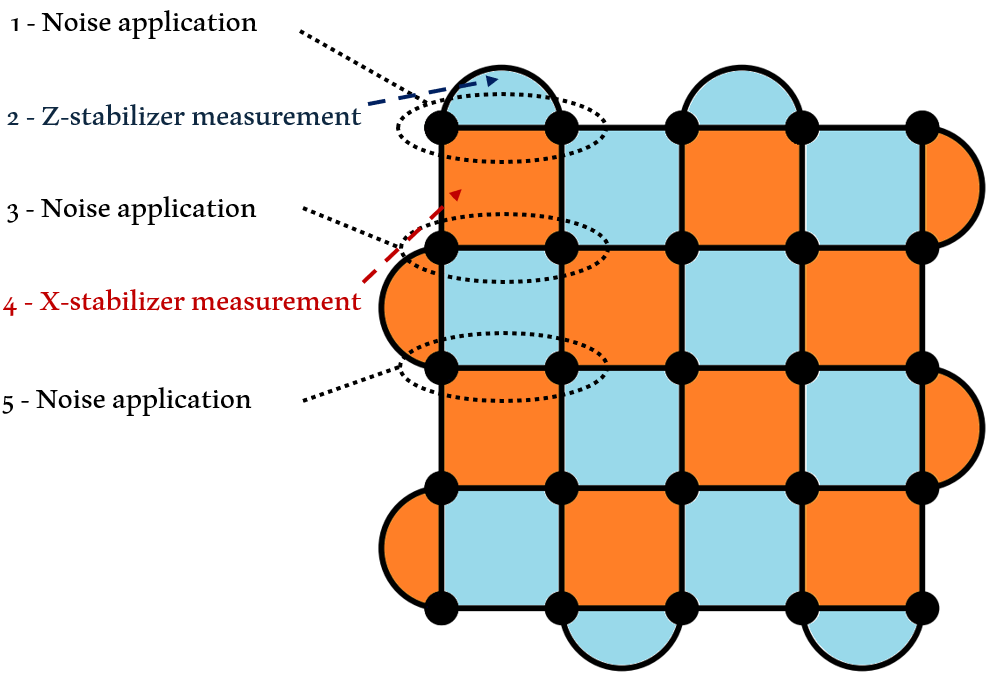}
  \caption{Layered noise application applied to simulation of phenomenological noise on rotated surface code.}
  \label{fig:layered_noise_phenom}
\end{figure}

Applying a non-Pauli noise channel such as amplitude damping or coherent unitary rotation to a qubit in our sparse representation typically doubles the number of populated basis components, since any Kraus operator that is drawn is a linear combination of two Pauli strings (see previous section). If noise were applied simultaneously to all $d^2$ data qubits, the number of populated basis labels might grow up to $2^{d^2}$, destroying sparsity and making simulation intractable. 

To mitigate this exponential blow-up, we exploit the complementary effect of projective stabilizer measurements, which tend to reduce the vector size by projecting the state back into one of the stabilizer eigenspaces. We therefore alternate noise application and stabilizer measurements in a carefully chosen order, ensuring that each region of the code is measured as soon as all its qubits have experienced noise (see Fig.~ \ref{fig:layered_noise_phenom}). This ``layered'' approach allows us to simulate phenomenological-level noise exactly (since the measurement of a stabilizer commutes with noise channels on qubits outside of the support of said stabilizer), while keeping the state-vector size under control throughout the evolution.

\paragraph{Noise layering.}
Starting from one corner of the rotated surface code, noise channels are applied incrementally to small subsets of qubits, followed immediately by the measurements of any stabilizers fully supported on those qubits. For example, consider the first few qubits of the lattice and their neighboring stabilizers:
\[
\begin{aligned}
Z_0Z_1,\;
& X_0X_1X_5X_6,\; \\
Z_1Z_2Z_6Z_7,\;
& X_5X_{10},\; \\
Z_5Z_6Z_{10}Z_{11},\;
& X_6X_7X_{11}X_{12}.
\end{aligned}
\]
A possible sequence (illustrated in  Fig.~\ref{fig:layered_noise_phenom}) proceeds as follows:
\begin{enumerate}
    \item Apply noise to qubits $0$ and $1$.
    \item Measure $Z_0Z_1$, since all its qubits have now been ``noisified.'' This measurement halves the number of populated components.
    \item To measure $X_0X_1X_5X_6$, noise must first be applied to qubits $5$ and $6$.
    \item Once qubits $5$ and $6$ have been updated, measure $X_0X_1X_5X_6$.
    \item Continue in this greedy fashion: for each stabilizer, apply noise to all its qubits if not already done, then measure it immediately once all its qubits have undergone noise.
\end{enumerate}

This procedure effectively sweeps across the lattice, alternating between layers of noise application and layers of stabilizer measurement. At any given moment, only the qubits belonging to a local patch of active stabilizers contribute to vector branching, while completed stabilizer regions are projected and compressed back into lower-rank subspaces.

\begin{figure}[ht]
    \centering
    \includegraphics[width=0.95\linewidth]{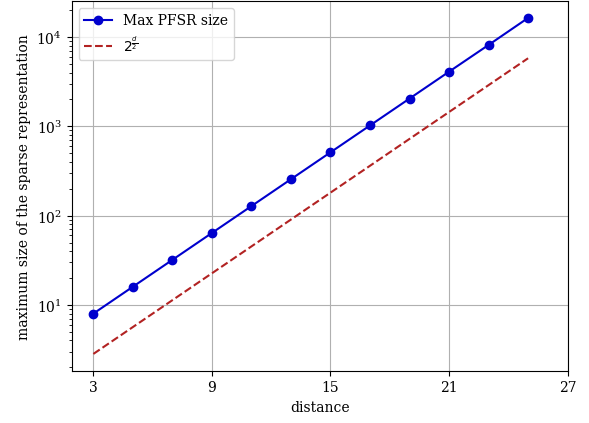}
    \caption{Maximal number of populated basis kets reached through the simulation of d+1 rounds of quantum error correction using the Pauli Frame Sparse Representation. Computed for a patch of rotated surface code submitted to phenomenological-level amplitude damping noise with $\gamma = 0.15$, under the layered noise application scheme.}
    \label{fig:size_evolution_phenomenological}
\end{figure}

Because of the locality and checkerboard structure of the rotated surface code, this layered schedule bounds the growth of the sparse vector to approximately $O(2^d)$ components, where $d$ is the code distance. Intuitively, at most one ``front'' of width proportional to $d$ (corresponding to the active column of stabilizers adjacent to the current measurement layer) remains unprojected at any given time. This significantly reduces both memory usage and runtime compared to a naive global noise application. In practice, this upper bound is seldom reached: in the case of amplitude damping or coherent noise, an error will usually only flip one type of stabilizer, either X- or Z-type, so the size of the sparse vector scales rather as $O(2^\frac{d}{2})$, as shown in Figure~\ref{fig:size_evolution_phenomenological}

Since the phenomenological noise model applies independent physical channels to each qubit, the layered strategy introduces no approximation: the order of noise and stabilizer measurements is irrelevant to the overall channel but strongly affects the intermediate sparsity of the simulated state. The same scheduling idea can be adapted to other lattice geometries or error-correction codes by following their stabilizer connectivity graphs and applying noise in local clusters before each stabilizer measurement layer, although codes with higher connectivities will see less computational benefits from it.

\subsubsection{Amplitude damping noise}

In this part, we compute a phenomenological-level threshold under the amplitude damping noise as described in Sec.~\ref{subsubsec:amplitude_damping}. We do so by computing the logical error rate after $d$ rounds of quantum error correction. We add measurement noise by flipping the measurement result with probability $\gamma$. For comparison, we also consider the Pauli Twirling Approximation (PTA)~\cite{Wallman2016PTA} of the amplitude damping channel. The PTA replaces the non-Pauli map $\mathcal{E}_{\mathrm{AD}}$ by a Pauli channel $\mathcal{E}_{\mathrm{PTA}}$ that reproduces the same action on the Pauli basis after random twirling. Explicitly,
\begin{equation}
\mathcal{E}_{\mathrm{PTA}}
= (1 - p_X - p_Y - p_Z)\, \mathbf{I}
  + p_X\, \mathbf{X} + p_Y\, \mathbf{Y} + p_Z\, \mathbf{Z},
\end{equation}
with probabilities obtained by projecting $\mathcal{E}_{\mathrm{AD}}$ onto the Pauli basis:
\begin{equation}
p_X = p_Y = \frac{\gamma}{4}, 
\qquad
p_Z = \frac{1 - \sqrt{1-\gamma}}{2} - \frac{\gamma}{4}.
\label{eq:pta-probabilities}
\end{equation}
This map reproduces the same average fidelity as the exact amplitude-damping channel to first order in $\gamma$ but eliminates all coherent terms. It is therefore efficiently simulatable by any stabilizer-based simulator such as Stim~\cite{Gidney2021stimfaststabilizer}.

\begin{figure}[ht]
    \centering
    \includegraphics[width=0.95\linewidth]{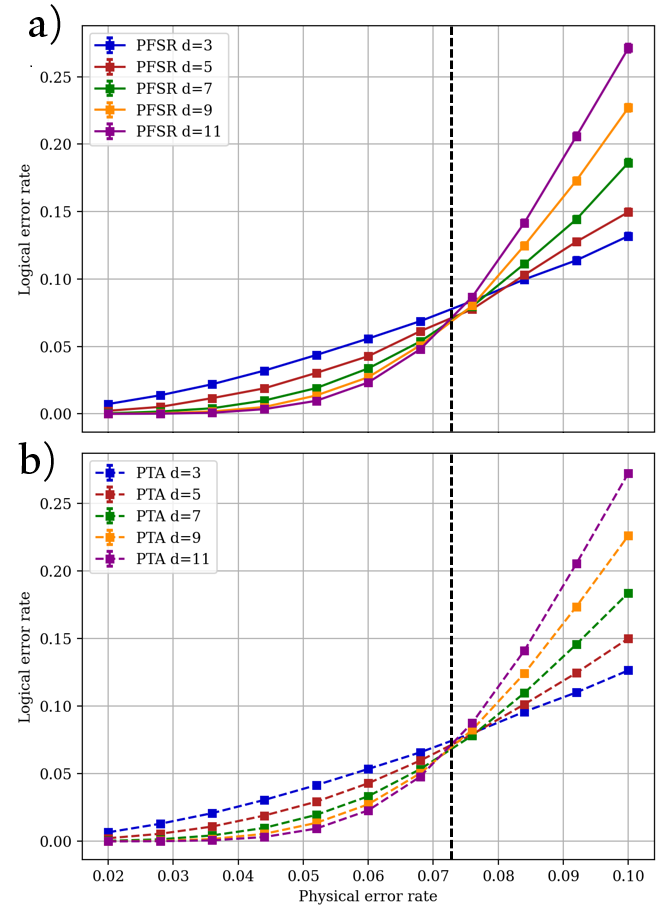}
    \caption{Logical error rate for the rotated surface code against amplitude damping noise at the phenomenological level. a) shows exact simulation of amplitude damping using the Pauli Frame Sparse Representation, while b) corresponds to the Pauli-twirled approximation simulated with Stim. Each point corresponds to $10^5$ trajectores in a) and $10^6$ trajectories in b). Both families of simulation show similar thresholds (indicated by the black dashed lines) $\gamma_\mathrm{exact} \approx \gamma_\mathrm{PTA} \approx 0.072$.}
    \label{fig:ampdamp_phenom_threshold}
\end{figure}

Figure~\ref{fig:ampdamp_phenom_threshold} shows the logical error rate after $d$ rounds of QEC as a function of the physical damping probability~$\gamma$ for rotated surface codes of distances $d=3,5,7,9,11$. Strikingly, the curves are nearly indistinguishable within statistical error: the extracted threshold values agree to within numerical precision: we find $p_\mathrm{th} \approx 0.72$. This indicates that, for amplitude-damping noise, the Pauli-twirled channel provides an exceptionally accurate effective description of logical behavior, at least at the phenomenological level.

\subsubsection{Coherent noise}

As another example of non-Pauli noise, we also compute a phenomenological-level threshold under the coherent noise described in Sec.~\ref{subsubsec:coherent_noise}. We compute the logical error rate after $d$ rounds of quantum error correction, and measurement noise is added by flipping the measurement result with probability $\sin{(\frac{\theta}{2})^2}$. We choose this value for measurement noise as it corresponds to the phase-flip probability of the PTA of the coherent noise channel,
\begin{equation}
\mathcal{E}_{\mathrm{PTA}}
= (1 - p_Z)\, \mathbf{I} + p_Z\, \mathbf{Z},
\end{equation}
with $p_z = \sin{(\frac{\theta}{2})^2}$.

A coherent Z-rotation error $U = e^{-i\frac{\theta}{2}} Z$ like the one we use has a trace-infidelity that scales as $1 - F \approx \theta^2$, but its action on off-diagaonal density-matrix elements is linear in $\theta$. In contrast, the PTA replaces the coherent rotation by a probabilistic Z-flip with probability $p_z = \frac{\theta^2}{4} + O(\theta^4)$, whose effect on the state is therefore quadratic in $\theta$ at all orders.Therefore, by cutting these first order off-diagonal terms, the PTA channel misses important information on phase coherence, and we expect the thresholds or logical-rate estimates based on the PTA to differ noticeably from those obtained under the exact coherent model.

\begin{figure}[ht]
    \centering
    \includegraphics[width=0.95\linewidth]{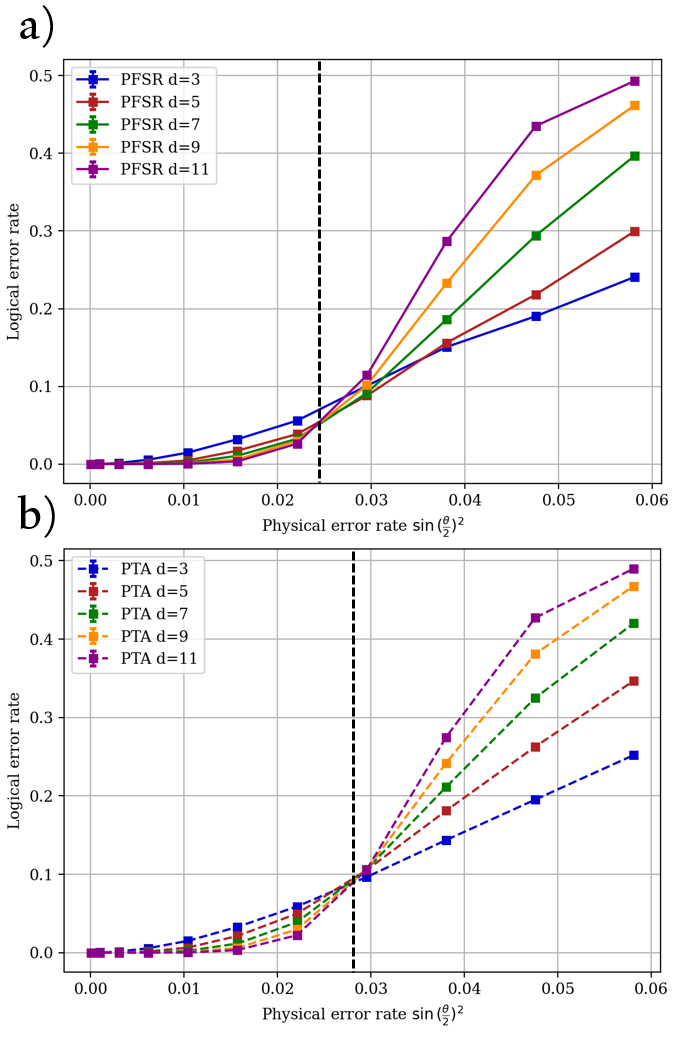}
    \caption{Same as Fig.~\ref{fig:ampdamp_phenom_threshold}, but for coherent noise $R_Z(\theta)$ (phenomenological level). The x-axis shows the physical error rate $\sin{(\frac{\theta}{2})^2}$ which is the phase-flip probability in the PTA simulation. Thresholds found for both simulations are $t_{\textrm{exact}} \approx 0.024$ and $t_{\textrm{PTA}} \approx 0.028$.}
    \label{fig:coherent_phenom_threshold}
\end{figure}

Figure~\ref{fig:coherent_phenom_threshold} shows the logical error rate after $d$ rounds of QEC as a function of the physical error rate~$\sin{(\frac{\theta}{2})^2}$ for rotated surface codes of distances $d=3,5,7,9,11$. The threshold observed at a physical error rate of $t_{\textrm{exact}} \approx 0.024$ is very close to the one obtained under the same conditions in Figure 3 of~\cite{Marton2023coherenterrors}. A slight difference of 0.004 is observed between the exact phenomenological threshold of coherent noise and the threshold obtained via PTA. This shows that Pauli-twirling is a perfectly fine approximation to compute a threshold at a qualitative level. However, we suspect that the situation might get worse once circuit-level noise is considered. This is what we investigate in the next subsection.

\subsection{Computing thresholds with circuit-level noise}

In order to accurately estimate quantum error correction thresholds, it is necessary to simulate down to the reality of hardware implementation. At the circuit level, we explicitly simulate the syndrome-extraction circuits, with ancilla preparation, entangling gates and measurement. Each elementary operation is subject to a noise process: single- and two-qubit gates undergo local error channels, measurements have finite fidelity, and ancilla qubits also undergo idling noise. Hence, this model includes the propagation of faults through two-qubit gates, which can in principle produce correlated (‘hook’) data errors. However, for the rotated surface code we adopt the standard CNOT orderings that prevent hook errors from reducing the effective code distance~\cite{Folwer2012SurfaceCode}. In our simulation, each gate is followed by the chosen physical noise channel (depolarizing or non-Pauli channel), while measurement results can be randomly flipped. Ancilla qubits are also subject to state-preparation error after each reset.

\subsubsection{Layered approach at the circuit level}

Performing fully parallel, optimal syndrome extraction (i.e., applying noise broadly and executing many ancilla-data CNOTs concurrently) causes a large temporary expansion of the sparse state in our Pauli Frame Sparse Representation: simultaneously applying non-Clifford noise to many data qubits increases branching exponentially with the number of affected qubits. To mitigate this blow-up we employ a layered circuit-level schedule, in which ancilla syndrome extraction is carried out sequentially (one ancilla at a time, executing the full syndrome extraction sequence: preparation, CNOTs and optional Hadamards, measurement and reset) according to the same locality-aware ordering we used for out phenomenological noise models in Sec.~\ref{subsubsec:layered_noise}. 

\begin{figure}[ht]
    \centering
    \includegraphics[width=0.95\linewidth]{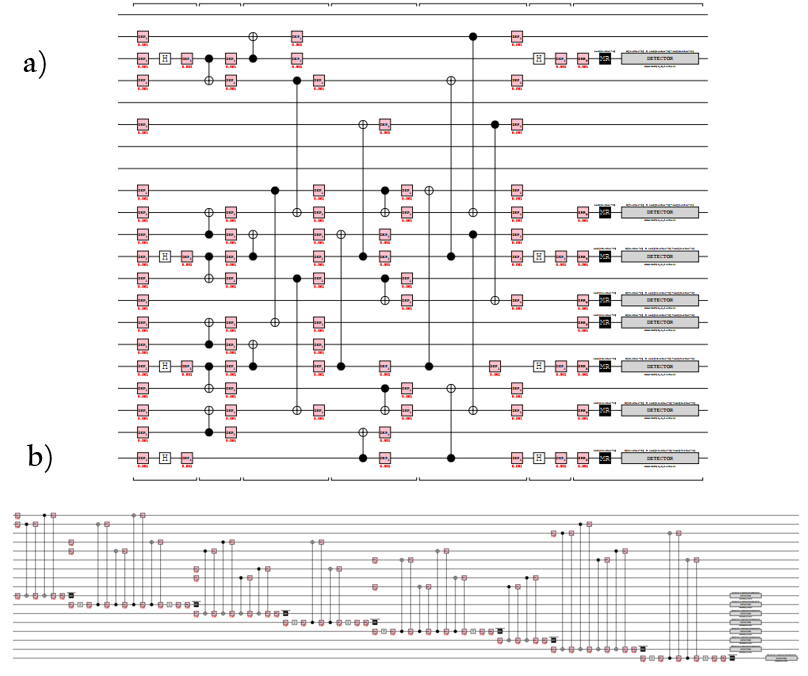}
    \caption{Circuit representation of one round of QEC, for a) optimal syndrome extraction, where CNOTs are applied in parallel for different stabilizers and b) layered syndrome extraction, where stabilizers are measured sequentially one by one. The pink boxes represent noise application, the black boxes are measurement of an ancilla. The stim circuit represented here is the one that was used for comparison purpose in Figure~\ref{fig:compare_parallel_layered}.}
    \label{fig:optimal_vs_layered}
\end{figure}

Concretely, for a target ancilla we (i) apply noise only to the data qubits required by its stabilizer (and to the ancilla), (ii) perform the ancilla’s full extraction circuit, and (iii) immediately perform the measurement (and the apply the projection to the sparse state) before moving to the next ancilla. The diagram of corresponding stim circuits are shown in Figure~\ref{fig:optimal_vs_layered}.

This sequential extraction schedule is suboptimal from a pure threshold point of view (parallel extraction minimizes the time that errors can spread before being measured), but it drastically reduces transient branching by ensuring that noise is applied only where necessary and that projective compression follows quickly. The schedule therefore makes circuit-level non-Pauli simulation tractable for larger distances than would be possible with fully parallel extraction.

\begin{figure}[ht]
    \centering
    \includegraphics[width=0.95\linewidth]{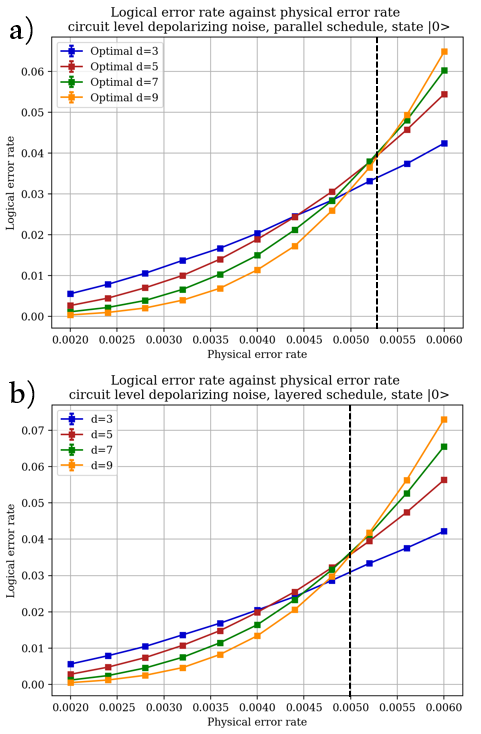}
    \caption{Logical error rate for the rotated surface code against circuit-level depolarizing noise. a) is obtained using the usual, optimal parallel scheduling of syndrome extraction. b) is obtained using the slightly suboptimal layered scheduling. The black vertical dashed curve corresponds to the estimate code thresholds $t_{\mathrm{parallel}} \approx 0.0053$ and $t_{\mathrm{layered}} \approx 0.0050$.}
    \label{fig:compare_parallel_layered}
\end{figure}

To validate that the sequential (layered) schedule is an acceptable approximation for threshold estimation we compare thresholds obtained with Stim under a standard depolarizing circuit-level noise model for two extraction schedules: (i) the usual parallel (optimal) schedule, and (ii) our layered/sequential schedule that measures ancillas one-by-one following the layered ordering. In Fig.~\ref{fig:compare_parallel_layered} we find that the extracted thresholds are very close: for the depolarizing model the threshold is approximately $t_{\mathrm{parallel}} \approx 0.0053$ for the parallel schedule versus $t_{\mathrm{layered}} \approx 0.0050$ for the layered schedule. This small difference ($3\times 10^{-4}$) indicates that the layered schedule does not substantially bias threshold estimation for the noise regimes we study, while providing large computational benefits for exact non-Pauli simulation.

Additionally, sequential or semi-serial syndrome extraction is not purely a numerical convenience: several physical architectures and compilation proposals naturally lead to serial or partially serial stabilizer readout \cite{Proctor2014SingleAncilla, Antipov2023SingleAncilla}. For example, architectures with global all-to-all connectivity (or with dynamic routing) such as neutral-atom~\cite{Muniz2025NeutralAtom} and trapped-ion~\cite{Bermudez2017IonsSingleAncilla, Ye2025IonsSingleAncilla} platforms increasingly consider ancilla reuse, mid-circuit measurement, moving ancillas, and sequential extraction as practical strategies.

The layered, sequential schedule therefore serves two purposes: (i) it facilitates simulation of circuit-level, non-Pauli noise using the Pauli Frame Sparse Representation, and (ii) it models a realistic operating point for architectures or compilation strategies where ancilla reuse and sequential extraction are natural or required. Because the threshold difference with the fully parallel schedule is small in our comparison, we regard the layered circuit-level thresholds reported below as representative of circuit-level performance for the considered noise models.

As amplitude damping noise seems to be particularly well approximated by PTA, we decided to focus our efforts on the circuit level towards the simulation of coherent noise.

\subsubsection{Truncation of small-amplitude terms}
\label{sec:truncation}

At the circuit level, even with the layered scheduling described above, exact simulation of non-Pauli noise channels and syndrome extraction will produce fault propagation that causes the sparse expansion to grow beyond practical memory limits. To keep simulation tractable, we introduce a controlled truncation step: after selected operations (in practice after any operation that increases support , i.e. non-Clifford gates and non-Pauli noise channels), we remove from the sparse vector any basis component whose amplitude absolute value falls below a fixed threshold $\varepsilon > 0$ and renormalize the retained state. This truncation process is similar in spirit to what is done in Pauli propagation methods~\cite{Begusic2024PauliPropagation, rudolph2025paulipropagationcomputationalframework}, whose aim is to keep a sparse representation of the observable.

If we represent the state in the current stabilizer frame as Eq.~\eqref{eq:pfsr_psi},
for a given cutoff $\varepsilon > 0$, we define the set of retained indices as 
\begin{equation}
    \mathcal{I}_{\geq \varepsilon} = \{ s\in\mathcal I \;:\; |\alpha_s| \geq \varepsilon \}.
\end{equation}
The truncation produces the 
normalized post-truncation state 
\begin{equation}
    \ket{\Psi'} =  \sum_{s\in\mathcal{I}_{\geq\varepsilon}} \frac{\alpha_s}{\sqrt{\nu}} \, P_s \ket{\mathbf{0}}
\end{equation}
where $\nu = \sum_{s\in\mathcal{I}_{\geq \varepsilon}} |\alpha_s|^2$.

\begin{figure}[ht]
    \centering
    \includegraphics[width=0.95\linewidth]{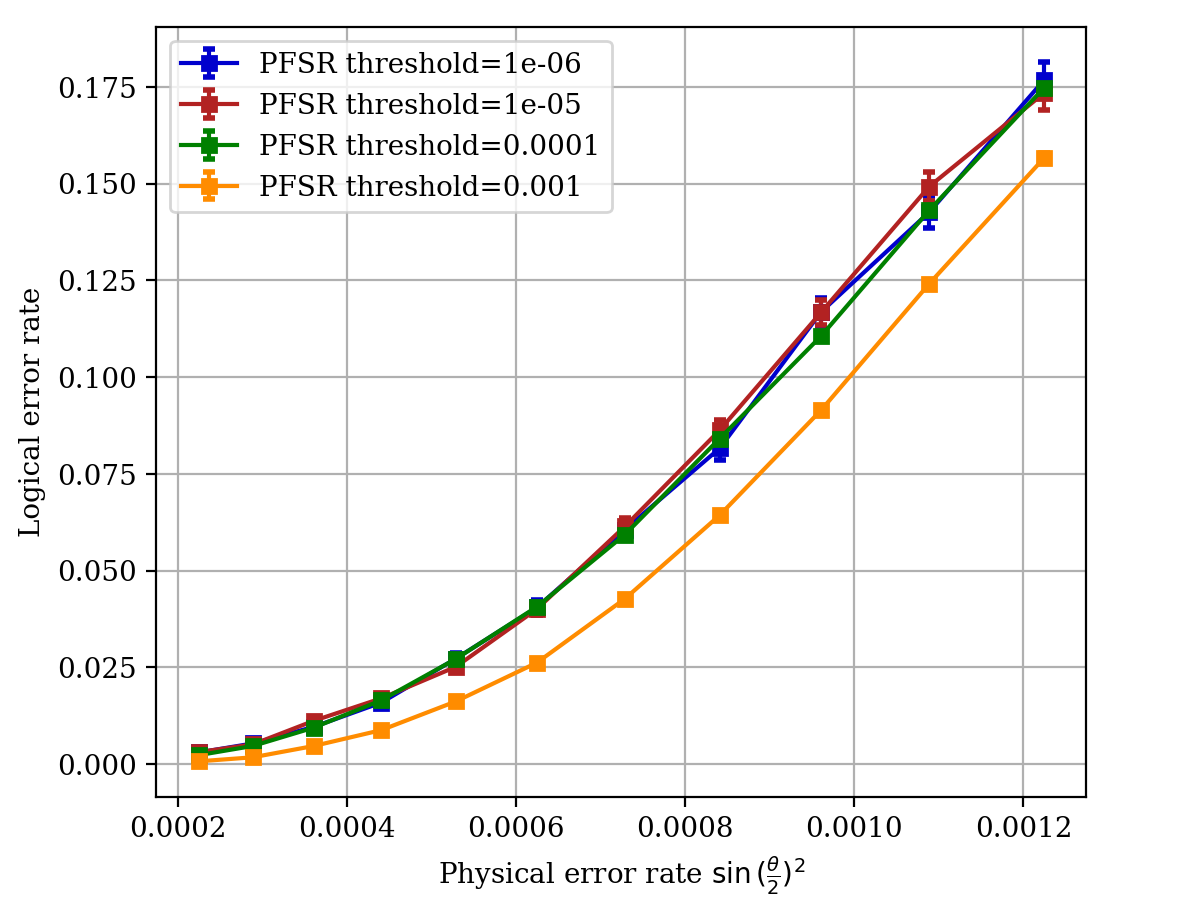}
    \caption{Logical error rate for the rotated surface code against circuit-level coherent noise, with different truncation cutoffs $\varepsilon$. 
    Logical error rate computed on 6 rounds of error correction on the logical $\ket{+}_L$ state.}
    \label{fig:compare_truncation_thresholds}
\end{figure}

We choose $\varepsilon$ by empirical convergence testing. Specifically, for distance $d=5$ we computed logical error vs physical error curves for a range of truncation cutoffs $\varepsilon$. As shown in Fig.~\ref{fig:compare_truncation_thresholds} the resulting logical-error curves closely overlap for a wide window of $0 < \varepsilon < 10^{-4}$, indicating that the truncation has negligible effect on the extracted threshold in that regime.

\begin{figure}[ht]
    \centering
    \includegraphics[width=0.95\linewidth]{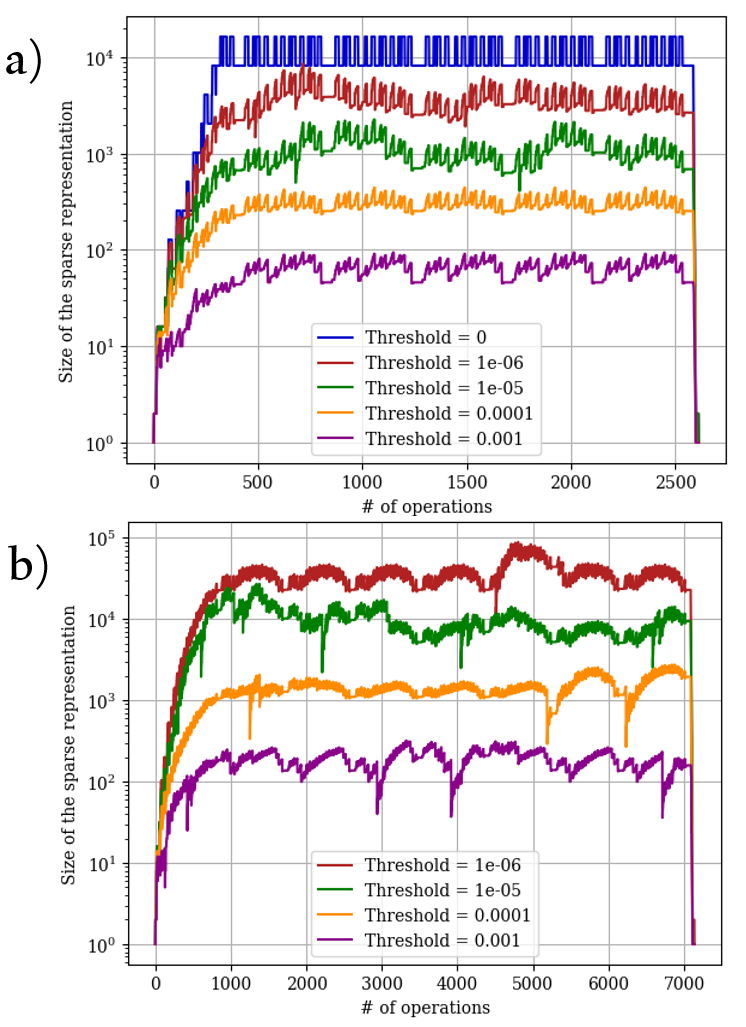}
    \caption{Evolution of the number of populated basis kets in the Pauli frame sparse representation through a) 6 rounds of error correction for the rotated surface code at $d=5$ b) 8 rounds of error correction for the rotated surface code at $d=7$.}
    \label{fig:size_evolution_truncation}
\end{figure}

As shown in Fig.~\ref{fig:size_evolution_truncation}, adding this truncation can reduce the number of populated basis kets by several orders of magnitude while retaining decent accuracy on the estimation of the error rate.

\subsubsection{Coherent noise}

\begin{figure}[ht]
    \centering
    \includegraphics[width=0.95\linewidth]{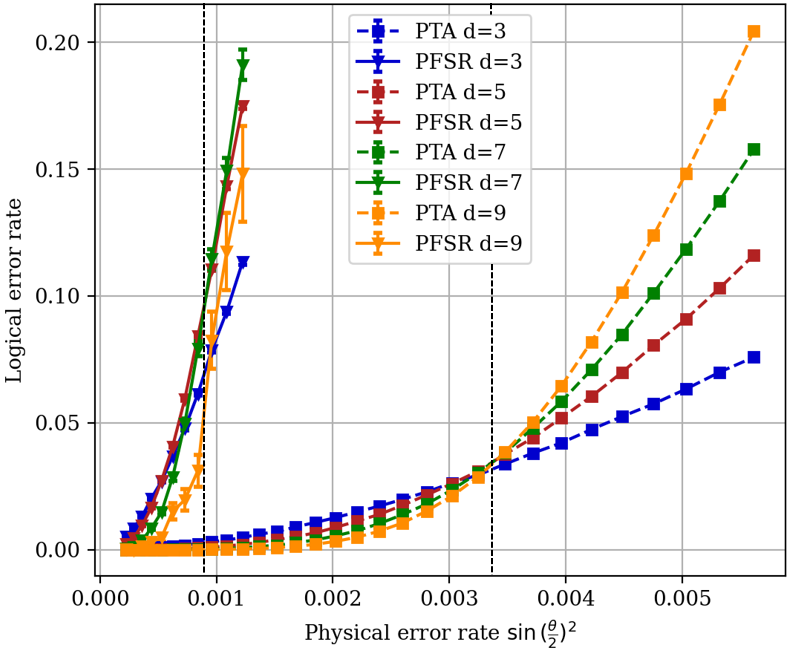}
    \caption{Logical error rate for the rotated surface code against unitary coherent noise $R_Z(\theta)$ at the circuit level. Solid curves with triangles correspond to truncated simulation of coherent noise using the Pauli Frame Sparse Representation, while dashed curves with squares correspond to the Pauli-twirled approximation simulated with Stim. Thresholds are indicated by the vertical dashed lines and are found to be ${t_\textrm{PFSR}} \approx 0.0009$ and $t_{\textrm{PTA}} \approx 0.0034$.}
    \label{fig:circuit_level_threshold_coherent}
\end{figure}

Figure~\ref{fig:circuit_level_threshold_coherent} presents the logical error rate as a function of physical coherent error strength for rotated-surface-code memory experiments at circuit level, for distances $d=3, 5, 7,$ and $9$. Two simulation methods are compared:
\begin{itemize}
    \item Pauli Twirling Approximation (PTA) simulation: the coherent noise is replaced by an equivalent stochastic Pauli channel and simulated using Stim.
    \item Pauli Frame Sparse Representation (PFSR) simulation: the evolution under coherent noise is tracked in the sparse expanded basis, with truncation cutoff $\varepsilon = 10^{-4}$.
\end{itemize}

The results show a striking qualitative difference. Whereas PTA predicts a threshold at a physical error rate of approximately ${t_\textrm{PFSR}} \approx 0.0034$, the full PFSR simulation yields a significantly lower threshold ${t_\textrm{PFSR}} \approx 0.0009$, a reduction by almost a factor of 4.

This discrepancy persists across distances and is already visible at moderate code sizes $d=5$ and $d=7$, where truncation effects remain negligible. For $d=9$, the onset of deviation between PFSR curves and expected scaling behavior suggests that the truncation threshold begins to limit accuracy. Nonetheless, the qualitative trend remains consistent: PTA systematically overestimates error-correcting performance in the presence of coherent rotations.

\section{Simulating magic state cultivation}
\label{sec:cultivation}

Beyond threshold estimation for surface codes, the PFSR-based simulator is well suited for analyzing near-Clifford circuits that incorporate non-Clifford resources in a structured way. As a representative and practically relevant example, we consider the magic-state cultivation circuit introduced by Gidney in Ref.~\cite{gidney2024magicstatecultivationgrowing}. In that work, the logical error rate per accepted shot is estimated for the injection and cultivation procedure at a distance $d=3$, using both $T$ gate and $S$ gate implementations. Gidney conjectures that the logical error rate associated with $T$ gates is approximately twice that of $S$ gates, but notes that verifying this conjecture at larger distances (e.g., $d=5$) is computationally prohibitive due to the number of qubits and the non-Clifford gates. In this section, we leverage the efficiency of the PFSR-based simulator to extend this analysis to larger code distances and directly test this conjecture. 

\subsection{Injection and cultivation circuit}

We simulate the magic-state injection and cultivation circuits at distances $d=3$ and $d=5$, following the construction introduced by Gidney. The circuits include both the injection stage and the cultivation protocol, but not the escape stage. Compared to the reference implementation provided in Gidney’s code, two modifications are required. First, as noted in the erratum of Ref.~\cite{gidney2024magicstatecultivationgrowing}, the double-check stage at distance $d=5$ is no longer trivially transversal due to the non-transversality of the $H_{XY}$ operation. We therefore incorporate the corresponding Pauli corrections directly into the layers of $T$ and $T^\dagger$ gates in the double-check procedure. Second, the original implementation does not include real-time Pauli frame updates during the growth step to restore newly measured stabilizers to the $+1$ eigenvalue. In our simulations, these corrections are applied dynamically, yielding a fully fault-tolerant circuit-level model.

The cultivation circuit provides a particularly natural benchmark for the PFSR framework. Although it contains non-Clifford resources, the circuit remains predominantly near-Clifford: the system leaves the code space only briefly during the double-check stage, and quickly returns to a stabilizer-dominated description. In our simulations, the PFSR expansion remains extremely sparse for most of the circuit execution, with the number of Pauli-frame terms typically remaining at two and peaking at 1024 terms only between the two non-Clifford layers of the double-check stage at $d=5$. This structure makes the cultivation protocol an ideal stress test for near-Clifford simulation methods. Table~\ref{tab:cultivation_circuits} gives further information on the type and the number of gates and noisy locations in each circuit. 

\begin{table}[h!]
\centering
 \begin{tabular}{||c c  c||} 
 \hline
 Metric & $d=3$ & $d=5$ \\ [0.5ex] 
 \hline\hline
 Total qubits & 15 & 42  \\ 
 Total gates (incl. measurements and resets) & 137 & 741  \\
 Two-qubit gates & 81 & 477  \\
 $T / T^\dagger$ gates & 15 & 53  \\
 Measurements & 14 & 93  \\
 Resets & 27 & 118  \\
 Noise channels & 504 & 3471  \\ [1ex] 
 \hline
 \end{tabular}
 \caption{Statistics of the different types of gate in the cultivation circuits. Errors on measurements are not included in the noise channel counts, so the cultivation circuits have in total $518$ (for $d=3$) and $3564$ (for $d=5$) potentially faulty locations. Note that during Monte Carlo simulations, we perform an extra round of perfect measurement of the stabilizers and of the logical $H_{XY}$, which are not accounted for in the above table.}
\label{tab:cultivation_circuits}
\end{table}

\subsection{Importance sampling for Monte Carlo simulations}

At distance $d=5$ and for relevant noise levels (around $p=10^{-3}$), we expect logical error rates per accepted shot to be of the order of $10^{-9}$, while the discard rate can be near 90\%. A brute-force Monte Carlo approach would therefore require on the order of hundreds of billions of circuit executions per noise value to obtain statistically meaningful estimates, rendering direct simulation extremely resource-heavy. This motivates us to employ an importance-sampling strategy inspired by some previous works on quantum error correction, such as ~\cite{Heussen2024ImportanceSampling, myers2025simulatinggeneralnoisenearly} where sampling is done over the number of faults or~\cite{Iyer2018ImportanceSampling, Hakkaku2021ISSyndrome} where sampling is directly done over the possible syndromes. Importance sampling over fault subsets allows us to target the rare fault configurations that actually contribute to logical failures, reducing the required number of samples by several orders of magnitude and making circuit-level studies of cultivation protocols all the more efficient. 

In the injection and cultivation circuits, all sources of error are represented by a bit-fip channel, a phase-flip channel, a depolarizing channel or a noisy measurement. All those faulty locations event share the same physical error probability $p$. Hence, we enumerate the $n_L$ potentially faulty locations in the circuit, and for each of them, a fault happens with probability $p$, meaning that the probability of having exactly $k$ faults in a circuit execution is 
\begin{equation}
    P(k) = \binom{n_L}{k} (1-p)^{n-k} p^k
\end{equation}
and for any event (in our case a logical error happening with the shot being kept, but it could be something else like the shot being discarded), we have 
\begin{equation}
    p_{\mathrm{fail}} = \sum_{k=0}^{n_L} P(k) p_{\mathrm{fail}|k},
\end{equation}
where $p_{\mathrm{fail}|k}$ is the conditional probability of a logical failure given exactly $k$ faults.

In practice, we estimate the conditional probabilities $p_{\mathrm{fail}|k}$ by Monte Carlo sampling circuits with exactly $k$ injected faults. Let $N_k$ denote the number of samples taken at fault number $k$, and let $N_k^{(\mathrm{fail})}$ denote the number of samples in which a logical failure occurs. We then form the estimator
\begin{equation}
    \hat{p}_{\mathrm{fail}|k} = \frac{N_k^{(\mathrm{fail})}}{N_k}.
\end{equation}
The overall error probability is estimated as
\begin{equation}
    \hat{p}_{\mathrm{fail}} = \sum_{k=0}^{n_L} P(k)\,\hat{p}_{\mathrm{fail}|k},
\end{equation}
which is an unbiased estimator of $p_{\mathrm{fail}}$.

This procedure can be viewed as a form of subset importance sampling, where the sample space is partitioned according to the number of faults. Instead of drawing circuits from the physical distribution of faults, which would overwhelmingly produce low-weight fault configurations, we explicitly condition on a fixed number of faults and reweight by the binomial factor $P(k)$. This allows us to efficiently probe the rare, high-weight fault configurations that dominate logical failure events.

This approach offers several advantages. First, the quantities $p_{\mathrm{fail}|k}$ are independent of the physical noise rate $p$, allowing a single set of simulations to generate logical error-rate curves over a wide range of noise parameters. Second, for a distance-$d$ fault-tolerant protocol operating on post-selection, any set of fewer than $d$ faults is either detected and discarded or accepted without any logical errors.  Therefore, $p_{\mathrm{fail}|k}=0$ for $k<d$, and the logical error probability is supported only on subsets with $k\geq d$. This property further concentrates the effective probability mass and justifies focusing computational effort on a narrow range of $k$ values above $d$, as contributions from large $k$ are exponentially suppressed at low $p$. Third, while the variance of a brute-force Monte Carlo estimator is dominated by the rarity of logical failures, subset sampling decouples the rarity of faults from the rarity of logical failures conditioned on those faults. By allocating sampling effort to values of $k$ for which $P(k)p_{\mathrm{fail}|k}$ is non-negligible, the estimator variance given as 
\begin{equation}
    \textrm{Var}(\hat{p}_{\mathrm{fail}}) = \sum_{k \geq d} P(k)^2 \frac{p_{\mathrm{fail}|k}(1 - p_{\mathrm{fail}|k})}{N_k}
\end{equation}
is reduced by several orders of magnitude.

\begin{figure}[ht]
    \centering
    \includegraphics[width=0.95\linewidth]{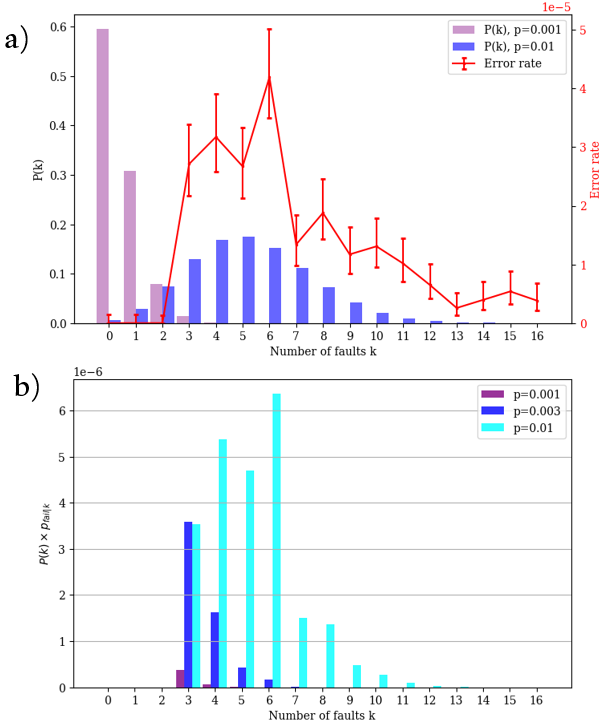}
    \caption{For a): the left axis shows distribution of probability of the number of faults in the cultivation circuit for $d=3$, for two values of noise $p=10^{-3}$ and $p=10^{-2}$. On the right axis: probability of getting a logical error while keeping the shot, conditioned by the number of faults $k$. Sub-figure b) shows the contribution $P(k) p_{\mathrm{fail}|k}$ to the total probability $p_{\mathrm{fail}}$ of keeping a shot and getting a logical error, for different noise values.}
    \label{fig:importancesamplingd3}
\end{figure}

These advantages are illustrated in Fig.~\ref{fig:importancesamplingd3}. Fig.~\ref{fig:importancesamplingd3}-a) shows both the probability distribution $P(k)$ of the number of faults and the logical failure rate per shot conditioned on said number of faults $p_{\mathrm{fail}|k}$. We can notice that as expected, no error can happen when we have $k < d$ faults in the circuit. This is due to the fault tolerant nature of the protocol, and means that no contribution to the variance will come from values $k < d$. Fig.~\ref{fig:importancesamplingd3}-b) shows the contribution of each number of faults $k$ to the total error rate, for different noise values $p$. At lower noise levels, the majority of the contributions come from the first few values of $k$ above $d$, and focusing our efforts on these most relevant numbers of faults will yield good results. From this graph, we can see that focusing on $3 \leq k \leq 16$ is more than enough.

In practice, we choose the number of samples $N_k$ adaptively, allocating more samples to values of $k$ for which $P(k)p_{\mathrm{fail}|k}$ contributes significantly to the total probability. For the $d=5$ cultivation circuit, this leads us to concentrate sampling effort on $k=5,6,7,8,9$, while higher-weight subsets contribute negligibly at the physical error rates of interest. This strategy yields accurate logical error estimates with computational costs several orders of magnitude lower than brute-force sampling. In contrast, estimating the discard rate requires sampling a broader range of $k$, although far fewer samples are needed per subset (a brute force approach is also perfectly fine to estimate the discard rate as it is note a rare event).

\subsection{Results}

Using this framework, we are able to extend Gidney’s original analysis to distance $d=5$ and directly test the conjectured relationship between $T$- and $S$-gate logical error rates.

\begin{figure}[ht]
    \centering
    \includegraphics[width=0.95\linewidth]{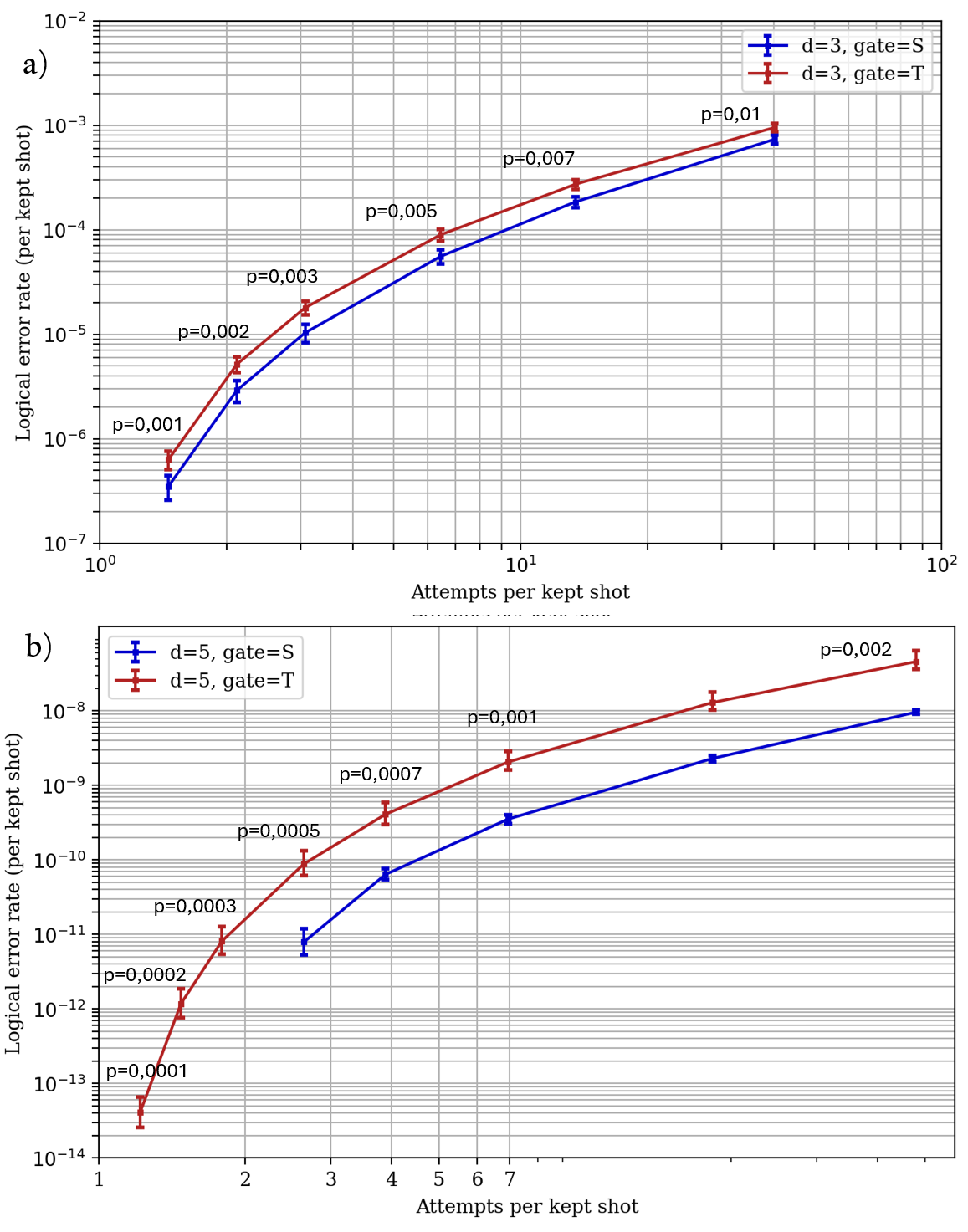}
    \caption{Logical error rate of the magic state cultivation protocol for $d=3$ in a) and $d=5$ in b), cultivating $T$ states (red) and $S$ states (blue). In both subfigures, the logical error rate was determined via importance sampling. In a), we sampled over all numbers of fault from $k=3$ to $k=16$, using around $7.5\times10^5$ shots per value of $k$, detecting between $3$ and $31$ logical errors depending on the value of $k$. In b), we sampled over $k = 5, 6, 7, 8, 9$, using between $10^9$ and $4\times10^9$ shots, detecting between $1$ and $4$ fault events for each value of $k$.}
    \label{fig:cultivation}
\end{figure}

In Fig.~\ref{fig:cultivation} we show the error rate of the magic state cultivation protocol, obtained via importance sampling on the number of faults. In particular, we can observe in Fig.~\ref{fig:cultivation}-b) that the discrepancy between the error rates of the $\ket{T}$ state and $S$ state injections become larger at $d=5$. We observe a factor as large as 7, compared to the factor 2 observed at $d=3$. 

We would like to highlight that this importance sampling approach becomes increasingly interesting the lower the physical error rate goes. Indeed, as the physical error rate diminished, the contributions to the logical error rate become increasingly dominated by the lower number of faults that can lead to logical failure, so in our case $k=d$. This allows us to compute the logical error rate for noises as low as we wish with only a few billion shots, while a brute force would require an unattainably high number of shots, for example more than $10^{13}$ at $p=10^{-4}$. 

During the preparation of this manuscript, we were made aware  of another work that aims to compute the logical error rate of the cultivation protocol at $d=5$~\cite{li2025softhighperformancesimulatoruniversal}, using the same type of sparse representation up to minor differences (such as keeping track of a destabilizer tableau in addition to the stabilizer tableau). We observe that the logical error rates we computed using importance sampling are in agreement with the brute-force calculations in Table 1 of~\cite{li2025softhighperformancesimulatoruniversal}, but with the main advantage that our importance sampling strategy requires only a few billions of shots for all noise values up to $p=2 \times 10^{-3}$, while \cite{li2025softhighperformancesimulatoruniversal} uses tens to hundreds of billions of shots per noise value.

\section{Conclusion}
\label{sec:conclusion}

In this work, we introduced the Pauli Frame Sparse Representation (PFSR) as a flexible and efficient tool for simulating near-Clifford quantum circuits under realistic noise models. The PFSR provides a compact representation that captures the action of general noise channels---including those far from Pauli or stochastic form---while remaining compatible with stabilizer-based simulation techniques. Because it preserves the structure of near-Clifford circuits without forcing a Pauli-twirl or a purely stochastic approximation, the PFSR bridges the gap between fully general noise descriptions and efficient classical simulation. This makes it well suited not only for modeling non-Pauli physical noise but also for analyzing circuits whose dynamics remain close to the Clifford stabilizer regime.

We demonstrated the usefulness of this framework through a detailed study of coherent error models, where the PFSR retains the leading-order coherent contributions that are suppressed by Pauli-twirled approximations. Using the rotated surface code as a testbed, we showed that this representation captures differences between the exact coherent channel and its twirled surrogate, particularly at small distances where the scaling mismatch between coherent and stochastic error components is most pronounced. While our phenomenological-level threshold estimates align with trends reported in prior work, our circuit-level investigation reveals a far more pronounced discrepancy: the exact coherent-noise threshold is nearly four times lower than the value predicted by the Pauli-twirled approximation, underscoring the importance of retaining non-Pauli structure when analyzing full fault-tolerant circuits.

Thanks to the PFSR, we were also able to compute the error thresholds of the recently introduced magic state cultivation protocol for up to distance $d=5$ with unprecedented shot efficiency. This allowed us to shed light on the quantitative different between the threshold obtained for $S$-state cultivation and for $T$-state cultivation.

\section{Acknowledgment}
We acknowledge useful discussions with Hui-Khoon Ng on Monte-Carlo simulations as well as with Hugo Jacinto about magic state cultivation.
TA is supported by France 2030 under the French National Research Agency award number
ANR-22-EXES-0013.
The simulations were executed on the Eviden Qaptiva platform.

\appendix

\section{Performing projections on anticommuting Pauli eigenspace}
\label{apdx:anticom_measurement}

\subsection{Computation of the Clifford $U$}

Let us consider a set of $n-1$ stabilizer generators $S_i$, for $0 \leq i \leq n-1$ (the $S'_{i\neq r}$ in the main text), and two other Paulis $A$ and $B$ (respectively $S_r$ and $P$ in the main text), all independent, such that all the $S_i$ commute with each other and with $A$ and $B$, and with $\{A, B\} = 0$. Our goal is to find a Clifford $U$ such that
\begin{equation}
    \begin{aligned}
        U S_i U^{\dagger} &= Z_i \\
        U B U^{\dagger} &= Z_n \\
        U A U^{\dagger} &= X_n.
    \end{aligned}
\end{equation}
We will do so by performing a Gaussian elimination on the symplectic representation of the Pauli operators, while also keeping track of the phases. We will write a Pauli $\sigma \in \mathcal{P}_n$ as 
\begin{equation}
    \sigma = i^q \prod_{k=1}^n Z^{z_k} X^{x_k}
\end{equation}
and represent it as a vector $(z_1, ... z_n, x_1, ..., x_n, q) \in \{0, 1 \}^{2n} \times \{0, 1, 2, 3\}$. We can then represent $(S_1, .., S_{n-1}, B, A)$ (in that order) by a matrix of size $(n+1) \times (2n+1)$, and our goal is to perform only Clifford operations to reduce it to the matrix
\begin{equation}
    \begin{bmatrix}
1 & 0 & 0 & \cdots & 0 & 0 & \cdots & 0 & 0 \\
0 & 1 & 0 & \cdots & 0 & 0 & \cdots & 0 & 0\\
\vdots & & \ddots & & \vdots & \vdots & & \vdots \\
0 & 0 & \cdots & 1 & 0 & 0 & \cdots & 0 & 0\\
0 & 0 & \cdots & 0 & 0 & 0 & \cdots & 1 & 0
\end{bmatrix}
\end{equation}

The algorithm is as follow:
\begin{enumerate}
    \item First, reduce the first $i \leq n$ rows to $Z_i$. For each row $R_i$, $i \leq n$ do the following:
    \begin{enumerate}
        \item Set a pivot $z_i = 1$. If $z_i = 0$, attempt the following in that order until one works:
        \begin{itemize}
            \item If $x_i = 1$, apply $H_i$
            \item If not, find a $j > i$ with $z_j = 1$, and apply $\textrm{CNOT}_{i \rightarrow j}$
        \end{itemize}
        \item Now that we have $z_i = 1$, use it to suppress all the other $z_j$ and $x_j$. For each $0 \leq j < n, j \neq i$, do in that order:
        \begin{itemize}
            \item if $z_j = 1$, apply $\textrm{CNOT}_{j \rightarrow i}$
            \item if $x_j = 1$, apply $H_j$ then $\textrm{CNOT}_{j \rightarrow i}$
        \end{itemize}
        \item Now that all $z_j = 0$ and $x_j = 0$, check if $x_i = 1$. If so, set it to zero by applying $H_i S_i H_i$
        \item We might still have a phase $q=2$, in this case, cancel it by applying $X_i$
    \end{enumerate}
    \item Now that the first $n$ rows are set to $Z_i$, we need to set the last row to $X_n$. 
    \begin{enumerate}
        \item Since our Paulis are all independent, the last qubit on the last row cannot be $I$. We then set our pivot $x_{n-1} = 1$ by doing the following:
        \begin{itemize}
            \item If $z_{n-1} = 1$ and $x_{n-1} = 1$, apply $S_{n-1}$
            \item If  $z_{n-1} = 1$ and $x_{n-1} = 0$, apply $H_{n-1}$
        \end{itemize}
        \item Then go over the other qubits $0 \leq j < n-1$, and get rid of the remaining $x_j = 1$ and $z_j = 1$ by doing in that order:
        \begin{itemize}
            \item if $x_j = 1$, apply $\textrm{CNOT}_{n-1 \rightarrow j}$
            \item if $z_j = 1$, apply $H_j$ $\textrm{CNOT}_{n-1 \rightarrow j}$ $H_j$ (this will not undo our efforts on the row $R_j$ above since conjugation by $H_j$ will turn $Z_j$ into $X_j$, so $\textrm{CNOT}_{n-1 \rightarrow j}$ will do nothing and the second Hadamard will send it back to $Z_j$)
        \end{itemize}
        \item Perform a last check to see if $z_{n-1} = 1$. If so, apply $S_{n-1}^{\dagger}$
        \item Finally, check the phase on the last row. If we still have a phase $q=2$, cancel it by applying $Z_{n-1}$
    \end{enumerate}
\end{enumerate}

This algorithm gives us the sequence of Clifford operation that composed together form the Clifford $U$ we are looking for.

\subsection{Update rule for Pauli history}

Once we are equipped with a Clifford $U$ performing the desired mapping, let us show how to update our Pauli histories. For a given Pauli history $P_{\mathbf{s}}$, our goal is to express the projected basis ket $\Pi_{\pm} P_{\mathbf{s}} \ket{\mathbf{0}}$ as a single Pauli history in the new stabilizer frame $\{S_1, ... , S_{n-1}, B\}$ (we can always reorder our stabilizers to have $B$ last). We have 

\begin{equation}
    \Pi_{\pm} P_{\mathbf{s}} \ket{\mathbf{0}} = \frac{I \pm B}{2} P_{\mathbf{s}} \ket{\mathbf{0}} = \frac{1}{2} P_{\mathbf{s}} \ket{\mathbf{0}} + \frac{1}{2} B P_{\mathbf{s}} \ket{\mathbf{0}}.
\end{equation}

Let us first look at what happens to $P_{\mathbf{s}} \ket{\mathbf{0}}$. We have 
\begin{equation}
    P'_{\mathbf{s}} = U P_{\mathbf{s}} U^{\dagger} = \underbrace{P_{\textrm{else}}}_{n-1 \,\ \text{qubits}} \otimes P_n
\end{equation}
and $U \ket{\mathbf{0}} = \ket{0}^{\otimes n - 1} \otimes \ket{+}$ in the computational basis, since $U$ maps the old stabilizer frame $\{S_1, ... , S_{n-1}, A\}$ to $\{Z_1, ... , Z_{n-1}, X_n\}$. Hence, we have
\begin{equation}
    \begin{aligned}
        P'_{\mathbf{s}} U \ket{\mathbf{0}} &= (P_{\textrm{else}} \otimes P_n) \ket{0}^{\otimes n - 1} \otimes \ket{+} \\
        &= (P_{\textrm{else}} \ket{0}^{\otimes n - 1}) \otimes (P_n \ket{+}) \\
        &= (P_{\textrm{else}} \ket{0}^{\otimes n - 1}) \otimes (\alpha \ket{0} + \beta \ket{1}) \\
        &= (P_{\textrm{else}} \ket{0}^{\otimes n - 1}) \otimes (\alpha \ket{0} + \beta X \ket{0}) \\
        &= [P_{\textrm{else}} \otimes (\alpha I + \beta X)] \ket{0}^{\otimes n} \\
        &= \alpha (P_{\textrm{else}} \otimes I) \ket{0}^{\otimes n} + \beta (P_{\textrm{else}} \otimes X) \ket{0}^{\otimes n}
    \end{aligned}
\end{equation}
so multiplying by $U^{\dagger}$ to the left gives (using the fact that $\ket{0}^{\otimes n} = U \ket{\mathbf{0}'}$ since $U$ maps the new stabilizer frame $\{S_1, ... , S_{n-1}, B\}$ to $\{Z_1, ... , Z_{n-1}, Z_n\}$):
\begin{equation}
    P_{\mathbf{s}} \ket{\mathbf{0}} = \alpha \underbrace{U^{\dagger} (P_{\textrm{else}} \otimes I) U}_{Q_1} \ket{\mathbf{0}'} + \beta \underbrace{U^{\dagger}(P_{\textrm{else}} \otimes X) U}_{Q_2} \ket{\mathbf{0}'}.
\end{equation}

Now what happens for the other part of the projector, $BP_{\mathbf{s}}$ ? We have 
\begin{equation}
    U B P_{\mathbf{s}} U^{\dagger} = U B U^{\dagger} U P_{\mathbf{s}} U^{\dagger} = Z_n P_{\mathbf{s}}' = P_{\textrm{else}} \otimes (Z_n P_n)
\end{equation}
so we get
\begin{equation}
    \begin{aligned}
        (BP_{\mathbf{s}})' U \ket{\mathbf{0}} &= (P_{\textrm{else}} \otimes (Z_n P_n)) \ket{0}^{\otimes n - 1} \otimes \ket{+} \\
        &= (P_{\textrm{else}} \ket{0}^{\otimes n - 1}) \otimes (Z_n P_n \ket{+}) \\
        &= (P_{\textrm{else}} \ket{0}^{\otimes n - 1}) \otimes (\alpha \ket{0} - \beta \ket{1}) \\
        &= \alpha (P_{\textrm{else}} \otimes I) \ket{0}^{\otimes n} - \beta (P_{\textrm{else}} \otimes X) \ket{0}^{\otimes n}
    \end{aligned}
\end{equation}
which gives
\begin{equation}
    B P_{\mathbf{s}} \ket{\mathbf{0}} = \alpha Q_1 \ket{\mathbf{0}'} - \beta Q_2 \ket{\mathbf{0}'}.
\end{equation}

Subsequently, when applying the full projector, we get 
\begin{equation}
    \Pi_{\pm} P_{\mathbf{s}} \ket{\mathbf{0}} = \begin{cases} \alpha Q_1 \,\ \textrm{for} \,\ \Pi_+ \\ \beta Q_2  \,\ \textrm{for} \,\ \Pi_- \end{cases}
\end{equation} 

showing the result of Eq.~(\ref{eq:history_anticom_proj}) of the main text.

\section{Stabilizer channel decomposition}
\label{apdx:stabilizer_channel_decomposition}

\subsection{Amplitude damping noise}

In its stabilizer channel decomposition~\cite{Bennink2017}, the amplitude damping noise channel is expressed as a linear combination of 3 channels:

\begin{equation}
    \mathcal{E} = q_I \mathbf{I} +  q_Z \mathbf{Z} + q_R \mathbf{R_z},
\end{equation}
where $q_I = \frac{(1-\gamma) + \sqrt{1 - \gamma}}{2}$, $q_Z = \frac{(1-\gamma) - \sqrt{1 - \gamma}}{2}$ and $q_R = \gamma$. $\mathbf{I}$ is the identity channel, $\mathbf{Z}$ is the $Z$ channel, and $\mathbf{R_z}$ is the reset channel, corresponding to a measurement of the qubit in the Z basis, and the application of an $X$ gate if the eigenvalue $-1$ is measured.

An advantage of this decomposition is that the channels $\mathbf{I}$ and $\mathbf{Z}$ are Paulis channels, and the reset is simply a measurement followed by a Pauli, so none of them will increase the number of populated basis kets of a PFSR.

However, this decomposition carries a significant drawback, as the coefficient $q_Z$ is negative, approximately $q_Z \approx -\frac{\gamma}{4}$ for $\gamma \ll 1$. This implies that the $q_i$s only form a quasiprobability distribution, with $q_I + q_Z + q_R = 1$. Hence, we will need to renormalize this quasiprobability distribution, to obtain probabilities between 0 and 1, defined as $p_i = \frac{|q_i|}{\sum_j |q_j|}$. Because of this renormalization, the statistical variance will increase exponentially with the negativity of our decomposition, which is the total magnitude of the negative coefficients, $\eta = \sum_{q_j < 0} |q_j|$. This implies that as we increase the number of noise channels we apply, we will need to exponentially increase the number of shots to maintain a constant variance. This method is then only viable for a relatively low number of amplitude damping noise channels with $\gamma << 1$. 

\subsection{Coherent noise}

Just like with amplitude damping noise, it is possible to use stabilizer channel decomposition to express this unitary channel as a linear combination of other channels that are easier to simulate. Here, the decomposition is 

\begin{equation}
    \mathcal{E} = q_I \mathbf{I} +  q_Z \mathbf{Z} + q_S \mathbf{S},
\end{equation}
where $q_I = \frac{1 + \cos{\theta} - \sin{\theta}}{2}$, $q_Z = \frac{1 - \cos{\theta} - \sin{\theta}}{2}$ and $q_S = \sin{\theta}$. $\mathbf{I}$ is the identity channel, $\mathbf{Z}$ is the $Z$ channel, and $\mathbf{S}$ is the S channel, which is simply the application of an S gate.

This method carries the same advantages and drawbacks as with amplitude damping noise, with all the channels being Clifford operations. However, the coefficient $q_Z$ is negative, approximately $q_Z \approx -\frac{\theta}{2}$ for $\theta << 1$, leading to the same exponential increase in the number of shots if we wish to maintain the variance under control.

\bibliographystyle{apsrev4-2}
\bibliography{bibliography.bib}

\end{document}